\documentclass[superscriptaddress,twocolumn,bibnotes,amsmath,amssymb,aps,pra,floatfix]{revtex4-2}

\usepackage{graphicx}
\usepackage{dcolumn}
\usepackage{bm}
\usepackage{hyperref}
\usepackage{physics}
\usepackage{braket}
\usepackage{multirow} 
\usepackage{xcolor}

\begin{document}

\title{Radio-Frequency Spectroscopy and the Dimensional Crossover in Interacting Spin-Polarized Fermi Gases}

\author{Jeff Maki}
\affiliation{Pitaevskii BEC Center, CNR-INO and Dipartimento di Fisica, Universit\`{a} di Trento, I-38123 Trento, Italy}
\author{Colin J. Dale}
\affiliation{Department of Physics and CQIQC, University of Toronto, Toronto, Ontario M5S 1A7, Canada}
\author{Joseph H.\ Thywissen}
\affiliation{Department of Physics and CQIQC, University of Toronto, Toronto, Ontario M5S 1A7, Canada}
\author{Shizhong Zhang}
\affiliation{Department of Physics and Hong Kong Institute of Quantum Science \& Technology, \\
The University of Hong Kong, Hong Kong, China}
\date{\today}

\begin{abstract} 
Low-dimensional ultracold gases are created in the laboratory by confining three-dimensional (3D) gases inside highly anisotropic trapping potentials. Such trap geometries not only provide access to simulating one-dimensional (1D) and two-dimensional (2D)  physics, but also can be used to study how the system crosses over towards a 3D system in the limit of weak confinement. In this work, we study the signature in radio-frequency (RF) spectroscopy for both the 1D-to-3D and the 2D-to-3D crossovers, in spin-polarized Fermi gases. We solve the two-body scattering T-matrix in the presence of strong harmonic confinement and use it to evaluate the two-body bound state and the RF spectroscopy transfer rate in the high frequency limit, covering both the quasi-low-dimensional and 3D limits. We find that in order to understand the dimensional crossover for spin-polarized Fermi gases with $p$-wave interactions, one needs to take into account an emergent $s$-wave interaction. 
\end{abstract}

\maketitle

\section{Introduction}

In recent years there has been an increased interest in understanding the physics of spin-polarized Fermi gases in low dimensions \cite{Cheon99, Gurarie05,Cheng06, Pricoupenko08, Gao15, Cui16a, Cui16b, Sekino18, Yin18, Fonta20, Tajima21, Maki21, Bertaina23,Jackson23, Marcum20, Chang20,Fedorov17,Levinsen08,Waseem17, Jiang18, Cui17, Zhang17}. These systems are special as they do not interact with the standard $s$-wave contact interaction, but rather a momentum dependent short-ranged $p$-wave interaction. In three-dimensional space (3D) such systems are plagued by severe three-body losses \cite{Suno03,Lasinio08, Yoshida15, Bertulani12, Waseem18,Schmidt20,Zhu22}, making the strongly interacting limit inaccessible. However, in low dimensions, it has been predicted that these systems are more stable than their 3D counterparts \cite{Cui16a, Cui16b, Cui17}. Moreover, strongly interacting low-dimensional spin-polarized Fermi gases offer interesting physics in their own right, such as topological superconductivity in two dimensions (2D) \cite{Fedorov17,Levinsen08, Gurarie05, Jiang18}, as well as integrable dynamics \cite{Cheon99} and the Bose-Fermi duality \cite{Cheon99, Valiente20a, Valiente20b} in one-dimensional space (1D). 

These low-dimensional spin-polarized Fermi gases can be readily accessed in the laboratory by tightly confining a 3D gas in an anisotropic external trapping potential. If the many-body energy scales, like temperature or chemical potential, are less than the energy gap to the first excited state of the trapping potential, all the atoms will reside in the the lowest lying state of the trapping potential; the system is effectively low-dimensional in this regime. One can then tune the effective one-dimensional interactions using the 3D $p$-wave Feshbach resonance, resulting in the confinement-induced resonances \cite{Cui17,Gao15, Pricoupenko08,Hess14, Saeidian15, Zhang17}.

On the other hand, when the energy scales of the many-body state become comparable to the energy spacing of the states of the trapping potential, the low-dimensional approximation begins to break down. The atoms can then begin to occupy higher states of the trapping potential. Once enough of these scattering continua are occupied, the system becomes fully 3D. Hence confined quantum gases are an excellent platform for studying not only low-dimensional physics, but also the dimensional crossover.

However, even when the low-dimensional energy hierarchy is intact, namely, when the relevant many-body energy scales are much smaller than the energy spacing of the trapping potential, 3D physics can still be revealed by probing the system at short distances. At scales much smaller than the transverse oscillator length, the confining potential is increasingly irrelevant. Hence tuning the length scale at which one probes the system can reveal  another type of dimensional crossover. 

There are several observables that may reveal this form of the dimensional crossover. For example, consider the particle distribution at high momentum: 
it is well known that the momentum distribution exhibits power-law tails for large momentum: $k \gg k_F$, where $k$ is the magnitude of the momentum and $k_F$ is the Fermi momentum \cite{Tan08a, Tan08b, Tan08c, Braaten08,Zhang09,Yu15,Cui16a, Cui16b,He21,Cayla23}. For $p$-wave gases, the high momentum tail changes continuously from $k_{\perp}^{-2}$ in the q2D limit ($k_{\perp} a_{\perp} \ll 1$) to $k^{-2}=(k_{\perp}^2 + k_{\parallel}^2)^{-1}$ in the 3D limit ($k_{\perp} a_{\perp} \gg 1$), where $k_{\perp (\parallel)}$ are the momentum components perpendicular (parallel) to the unconfined two-dimensional space \cite{Cui16a, Cui16b,He21}. The constants of proportionality for the high-momentum tails in the quasi-low- and high-dimensional limits are related to the $d$-dimensional thermodynamic contacts, which depend on the many-body details of the system and how many states of the confining potential are occupied.

In order to study the dimensional crossover using the momentum distribution, one needs the ability to measure the momentum distribution over many decades ranging from $k_{\perp} a_{\perp} \ll 1$ to $k_{\perp} a_{\perp} \gg 1$. 
Although this is possible using time-of-flight expansion,
one needs to optimize the free-expansion time to maintain a good signal-to-noise ratio over many orders of magnitude of momenta. Even in experiments well adapted to such a measurement, spurious $k^{-4}$ tails have been observed due to spin impurities \cite{Cayla23}. Moreover, the continuous nature of the momentum distribution makes it difficult to distinguish definitive signatures of the dimensional crossover.

A more commonly used technique to study the high-momentum distribution is radio-frequency (RF) spectroscopy \cite{Luciuk16, Tan08a, Tan08b,Tan08c, Braaten08, Braaten10, Son10, Baur12, Langmack12, Sagi12, Fischer14}. In RF spectroscopy, a radio-frequency pulse performs a spin flip from some initial interacting hyperfine state $|a\rangle$ to a final non-interacting hyperfine state $| b \rangle$ at a fixed detuning $\omega$. For large detunings (compared to the energy spacing between the hyperfine states) the RF transfer rate has power-law tails, and the constants of proportionality are again related to the thermodynamic contacts. The benefit of the RF spectroscopy transfer rate is that the dimensional crossover can be studied by simply adjusting the detuning, independent of whether the initial state is in the low-dimensional regime or not. 

In this work we examine the 1D-to-3D and 2D-to-3D crossovers in $p$-wave spin-polarized  Fermi gases using the RF spectroscopy transfer rate. We obtain analytical results for the two-body scattering in both quasi-one-dimensional (q1D) and quasi-two-dimensional geometries (q2D). This allows us to evaluate the two-body bound states, as well as the RF transfer rate at the two-body level for the whole dimensional crossover. The results for the RF transfer rate are made more rigorous using the operator product expansion (OPE) in the low-dimensional and three-dimensional limits, which can connect the RF tails to the corresponding thermodynamic contacts. Our results for the two-body bound states and the RF transfer rate indicate that there are two effective low-dimensional interactions, one that is $p$-wave and one that is $s$-wave. Both interactions are necessary for understanding the dimensional crossover, although only the low-dimensional $p$-wave interaction is needed for understanding the low-dimensional limit. 

Our study was motivated by a recent experiment \cite{Jackson23} 
that found evidence of q1D $s$-wave scattering in spin-polarized Fermi gas and a first signature of the dimensional crossover in RF spectroscopy. Here we provide a complete theoretical description of both the 1D-to-3D and the 2D-to-3D crossovers for the RF transfer rate and the dimer state. This work is also accompanied by an experimental work \cite{exp_companion} examining the RF spectroscopy and binding energies in q2D Fermi gases.

The remainder of this article is organized as follows. In Sec.~\ref{sec:twobody} we discuss how three-dimensional $p$-wave scattering in the presence of harmonic confinement can produce both low-dimensional $p$-wave and $s$-wave interactions. In Sec.~\ref{sec:q1D}, we then explicitly solve the Lippmann-Schwinger equations for two-body $p$-wave scattering in q1D systems. We then extend this formalism to q2D systems in Sec.~\ref{sec:q2D} and calculate the two-body bound states along the dimensional crossover in Sec.~\ref{sec:bound states}.
In Sec.~\ref{sec:RF} we determine the RF spectroscopy transfer rate for the whole dimensional crossover. The conclusions are presented in Sec.~\ref{sec:conc}.

\section{p-Wave Scattering in Harmonic Confinement}
\label{sec:twobody}

Consider two non-interacting identical fermions in an anisotropic external harmonic trap:
\begin{equation}
    U_{\text{ext.}}({\bf r})=\frac{1}{2}m(\omega_\perp^2 {\bf r}_\perp^2+\omega_\parallel^2 r_\parallel^2)\,,
\end{equation}
where ${\bf r}_\perp$ is the position vector in the confined directions (the transverse directions), while ${\bf r}_{\parallel}$ is for the weakly confined directions (the axial directions). To be specific, in the q1D case:  ${\bf r}_{\perp} =(x,y)$  and ${\bf r}_{\parallel} = z$, while for the q2D trap: ${\bf r}_{\perp} = z$ and ${\bf r}_{\parallel} =(x,y)$. In both cases, we consider high anisotropy and require the transverse oscillator frequency $\omega_\perp$ to be much larger than that of the axial directions $\omega_\perp\gg \omega_\parallel$. Furthermore, we shall set that the axial oscillator frequency $\omega_\parallel=0$ and assume that any effects with a non-zero $\omega_\parallel$ in actual experiment can be adequately taken care of using the local density approximation. 

At the two-body level, the center of mass motion can naturally be separated from the relative motion. While in general the center of mass motion of the pairs cannot be ignored, for the discussion of the leading high-frequency tails of the RF transfer rate, the inclusion of the center of mass is irrelevant. The relative motion is described by the simple Hamiltonian:
\begin{equation}
H = \frac{ {\bf p}_\parallel^2}{2\mu} + \frac{\bf p_{\perp}^2}{2\mu}  + \frac{1}{2}\mu \omega_{\perp}^2 {\bf r}_{\perp}^2 + v(r)\,,
\label{eq:2B_Hamiltonian}
\end{equation}
where $\mu$ is the reduced mass, ${\bf p}_{\parallel}$ is the axial momentum along the unconfined direction, and ${\bf p}_\perp$ is the momentum in the transverse directions. The interaction potential $v(r = |{\bf r}|)$ is short ranged; the range $r_0$ is much smaller than the typical size of the trapping potential. In particular $r_0\ll a_\perp$ where $a_\perp=\sqrt{1/\mu\omega_\perp}$ is the transverse oscillator length. 

To understand the effects of the short-range potential $v(r)$ on the scattering of two identical fermions, it is beneficial to examine its matrix element between two non-interacting states of the relative motion of the two fermions $\phi({\bf r})$ and $\chi({\bf r})$:
\begin{equation}
   \langle \phi | v(r) | \chi \rangle=\int d^3{\bf r} \phi^*({\bf r})v(r)\chi({\bf r})\,.
\end{equation}
Because of the fermionic statistics, the relative wavefunctions must be odd under exchange, or equivalently parity: $\phi({\bf r})=-\phi(-{\bf r})$ and $\chi({\bf r})=-\chi(-{\bf r})$.
Assuming that the scale over which $\phi({\bf r})$ and $\chi({\bf r})$ changes are much larger than $r_0$, i.e.\ if we look at states with energies much smaller than $r_0^{-2}$, we can then expand the integrand around ${\bf r}=0$ and obtain:
\begin{align}\nonumber
&\int d^3{\bf r} (\phi^*({\bf 0})+r_i\boldsymbol{\nabla}_i\phi^*({\bf 0}))v(r)(\chi({\bf 0})+r_j\boldsymbol{\nabla}_j\chi({\bf 0}))\,,\\\nonumber
=&\int d^3{\bf r} r_iv(r)r_j\boldsymbol{\nabla}_i\phi^*({\bf 0})\boldsymbol{\nabla}_j\chi({\bf 0})\,,\\\nonumber
=& \left[\int d^3{\bf r} x^2v(r)\right]\boldsymbol{\nabla}\phi^*({\bf 0})\cdot \boldsymbol{\nabla}\chi({\bf 0})\equiv g\boldsymbol{\nabla}\phi^*({\bf 0})\cdot \boldsymbol{\nabla}\chi({\bf 0})\,.
\end{align}
where we have used the obvious result that $\chi({\bf 0})=\phi({\bf 0})=0$ in the second line. In the last line, we also used the fact that integration such as $\int d^3{\bf r} xyv(r)$ are zero due to rotational symmetry. From the final equality, the p-wave matrix element at leading order is equivalent to that of a phenomenological $p$-wave pseudo-potential:
\begin{equation}
    v(r) = g \overleftarrow{\boldsymbol{\nabla}}_i \delta({\bf r})  \overrightarrow{\boldsymbol{\nabla}}_i \,, 
    \label{eq:pseudopotential}
\end{equation}
where the arrows indicate the direction to which the gradient is applied. This allows us to identify the bare coupling constant in terms of the microscopic two-body potential as: $g\equiv \int d^3{\bf r} x^2v(r)$. In the particular case when both the initial state and final state are described by plane waves with wave vector ${\bf k}$ and ${\bf k}'$, then clearly
\begin{equation}
    \langle {\bf k } | v(r) | {\bf k'} \rangle=\frac{g}{L^3}{\bf k }\cdot{\bf k }'\,,
\end{equation}
where $L^3$ is the volume of the system.

Once the form of the pseudo-potential Eq.(\ref{eq:pseudopotential}) is chosen, it is clear that the requirement for antisymmetrizing the non-interacting wave function $\phi$ and $\chi$ is no longer necessary since the even parity part is automatically zero in the transition matrix element. One can improve the above approximation by taking into account additional derivatives such as $\int d^3{\bf r} x^2y^2v(r)\boldsymbol{\nabla}^2_x\phi^*({\bf 0})\boldsymbol{\nabla}^2_y\chi({\bf 0})$ that are generally suppressed at low energy by the additional derivatives. 

For 3D $p$-wave scattering in the absence of the harmonic trap, the $p$-wave pseudo-potential in Eq.~(\ref{eq:pseudopotential}) suggests that the low-energy scattering $T$-matrix, $\hat{T}_p$, can be written in the following form at low-energy~\cite{Ahmed-Braun21}:
\begin{align}
\langle {\bf k } | T | {\bf k'} \rangle  &= 2{\bf k \cdot k'} T_{3D,p}(E)  \label{eq:3D_T_matrix1}\,,\\
T_{3D,p}^{-1}(E) &= \frac{2\mu}{24\pi}\left[ \frac{1}{V_p} + \frac{2\mu E}{R_p}  +(-2\mu E -i \delta)^{3/2}\right]\,,
\label{eq:3D_T_matrix}
\end{align}
where $|{\bf k^{(')}}\rangle $ describes a plane wave state with momentum ${\bf k^{(')}}$ and $|{\bf k}|  = |{\bf k'}|$. The factor of $2$ in the matrix element accounts for the indistinguishable nature of the fermions.  In Eq.~(\ref{eq:3D_T_matrix}), the $p$-wave scattering volume, $V_{p}$, and the 3D $p$-wave effective range, $R_{p}$, depend on the details of the potential $v(r)$~\cite{Ahmed-Braun21}, but such knowledge is not needed for our purposes. 

In the presence of the harmonic confinement, Eq.~(\ref{eq:pseudopotential}) is still valid since it only refers to the short-range form of the wave functions which is insensitive to external confinement. However, now the non-interacting states are plane waves with momentum ${\bf k}$ in the axial directions, while in the transverse direction they are harmonic oscillator states. The pseudo-potential can now act on either the axial or transverse part of the two-body wavefunction. If the gradients act on the axial directions, the matrix elements of the $T$-matrix will have the form of a low-dimensional $p$-wave interaction: $\langle {\bf k} | v| {\bf k}' \rangle \propto g{\bf k \cdot k'}$. This occurs when the axial two-body wavefunctions are odd functions under parity (or equivalently particle exchange), while the transverse two-body wavefunctions are even function under parity. This is the only possibility when all the atoms reside in the lowest state of the transverse trapping potential.

If instead the gradients act on the transverse directions one obtains an interaction reminiscent of a low-dimensional $s$-wave interaction: $\langle {\bf k} | v| {\bf k}' \rangle \propto g$. In order to have a non-zero matrix element, the two-body wavefunction has to have even parity along the axial direction and odd parity along the transverse direction. This can only be accomplished when the atoms occupy different single particle transverse states which allows for an odd parity state. 

In the following sections, we will specifically consider scattering in q1D and q2D geometries and show how starting from the fully 3D scattering, one reproduces the results of low-dimensional $s$- and $p$-wave scattering, before we investigate the crossover of radio-frequency spectral from its low-dimensional feature to that corresponds to full 3D as one increases the probing rf-frequency. 

\section{p-wave Scattering in q1D}
\label{sec:q1D}

We now turn to a specific calculation of the two-body scattering properties inside a q1D harmonic trapping potential. The non-interacting eigenstates and energies (in the center of mass frame) are given by :
\begin{align}
\psi_{k,n,m}(r_{\parallel},{\bf r_{\perp}}) &= e^{i k r_{\parallel}} \sqrt{\frac{n!}{\pi a_{\perp}^2(n+|m|)!}} \nonumber \\
&e^{-\frac{r_{\perp}^2}{2 a_{\perp}^2}} e^{i m \phi} \left(\frac{r_{\perp}}{a_{\perp}}\right)^{|m|} \mathcal{L}^{|m|}_n \left(\frac{r_{\perp}^2}{a_{\perp}^2}\right) \,,   \label{eq:q1d_basis} \\
E_{n,m}(k) &= \frac{k^2}{2\mu} + (2n + |m|+1)\omega_{\perp} \,.\label{eq:q1D_energy}
\end{align}
In Eqs.~(\ref{eq:q1d_basis}-\ref{eq:q1D_energy}) $k$ is the axial momentum, $n$ is the transverse principle quantum number, and $m$ is the angular momentum in the transverse direction. We have also written ${\bf r}_{\perp}  = (r_{\perp}, \phi)$ in polar coordinates. In this representation the transverse eigenstates have definite parity: $P = (-1)^m$.

The matrix elements of the pseudopotential Eq.~(\ref{eq:pseudopotential}) in this basis are found to be:
\begin{align}
    \langle k n m | &v | k' n' m \rangle = \nonumber \\
    &\frac{2g}{\pi a_{\perp}^2}\left( k k' \delta_{m,0} + \frac{2}{a_{\perp}^2} \sqrt{n+1}\sqrt{n'+1} \delta_{|m|,1}\right)\,.
    \label{eq:q1D_matrix_elements}
\end{align}
When $m=0$, the interactions are of 1D $p$-wave character with the required form factor $kk'$, while for $m = \pm 1$ the interactions resemble a 1D $s$-wave interaction. This follows from the parity considerations discussed in the previous section.

We now solve the Lippmann-Schwinger equation for the two-body $T$-matrix inside the harmonic confinement. Since the angular momentum $m$ is a conserved quantity, the different angular momentum sectors can be evaluated independently. The full details of the calculation are presented in Appendix~\ref{app:LSE}. In the case when $m=0$ which corresponds to $p$-wave scattering, the matrix element of $T$ can be written as
\begin{equation}
\langle k,n,0| T | k',n',0 \rangle= 2 k k' T_{p,\mathrm{1D}}(E)\,,
\end{equation}
with $T_{p,\mathrm{1D}}(E)$ given explicitly by
\begin{equation}
T_{p,\mathrm{1D}}^{-1}(E)= 
\frac{2\mu }{2}\left[\frac{a_{\perp}^2}{6} \left(\frac{1}{V_p} + \frac{2\mu E}{R_{p}}\right)  - \frac{2}{a_{\perp}} \zeta\left(-\frac{1}{2},-\mathcal{E}_0 - i \delta\right)\right]\,.
\label{eq:T_0}
\end{equation}
In Eq.~(\ref{eq:T_0}), $\mathcal{E}_m  = \left(E-(|m|+1)\omega_{\perp}\right) / 2\omega_{\perp} = \left(k a_{\perp}/2\right)^2$ is the ratio of the kinetic energy in the axial direction for the $m$th channel to the transverse level spacing $2\omega_{\perp}$, and $\zeta(s,x)$ is the Hurwitz zeta function. For the $m = \pm 1$ sector, we have similarly
\begin{align}
\langle k,n,\pm 1 | &T | k',n', \pm 1 \rangle = 2\sqrt{n+1}\sqrt{n'+1} T_{s,\mathrm{1D}}(E)\,, \nonumber \\
T_{s,\mathrm{1D}}^{-1}(E)&=
\frac{2\mu }{2}\left[\frac{a_{\perp}^4}{12} \left(\frac{1}{V_{p}} + \frac{2\mu E}{R_{p}}\right)  \right. \nonumber \\
+ \frac{a_{\perp}}{2} &\left.\left[ \zeta\left(-\frac{1}{2},-\mathcal{E}_1 - i \delta\right) + \frac{E}{2\omega_{\perp}}\zeta\left(\frac{1}{2},-\mathcal{E}_1 - i \delta\right)\right]\right]\,. \nonumber \\
\label{eq:T_1}
\end{align}
Eqs.~(\ref{eq:T_0}-\ref{eq:T_1}) are the exact solutions to the Lippman-Schwinger equation and are valid for arbitrary energies. 

The low-dimensional limit is defined when the kinetic energy is much less than $2\omega_{\perp}$, i.e. $\mathcal{E}_m \ll 1$. First consider $m=0$. In the 1D or low-energy limit, $\mathcal{E}_0 \ll 1$, Eq.~(\ref{eq:T_0}) reduces to:
\begin{align}
T^{-1}_{p,\mathrm{1D}}(E) &=\frac{2\mu }{2} \left[\frac{1}{a_{p,\mathrm{1D}}} + k^2 r_{p,\mathrm{1D}} +i k \right]\,, \label{eq:T_1D_p} \\
\frac{1}{a_{p,\mathrm{1D}}} &= \frac{a_{\perp}^2}{6}\left( \frac{1}{V_{p}} + \frac{2}{R_{p} a_{\perp}^2}\right) - \frac{2}{a_{\perp}} \zeta\left(-\frac{1}{2}\right)\,, \label{eq:1d_ap}\\
r_{p,\mathrm{1D}} &= \frac{a_{\perp}^2}{6R_{p}} + \frac{a_{\perp}}{4} \zeta\left(\frac{1}{2}\right)\,,
\label{eq:1d_rp}
\end{align} 

\noindent where $\zeta(s) = \zeta(s,1)$ is the Riemann zeta function. The low-energy form of the T-matrix is the exact result one would obtain from a true 1D calculation with $p$-wave interactions in 1D. Here $a_{p,\mathrm{1D}}$ and $r_{p,\mathrm{1D}}$ are the 1D $p$-wave scattering volume and effective range, respectively, and are written in terms of the 3D scattering parameters, $V_{p}$ and $R_{p}$. Eqs.~(\ref{eq:1d_ap}-\ref{eq:1d_rp}) have been reported previously \cite{Cui17,Gao15, Pricoupenko08,Hess14, Saeidian15}. 

Similarly for small kinetic energies in the even-wave sector, $\mathcal{E}_1 \ll 1$, the $m=\pm 1$ T-matrix reduces to:

\begin{align}
T_{s,\mathrm{1D}}^{-1}(E) =& \frac{2\mu }{2} \left[-a_{s,\mathrm{1D}} + k^2 r_{s,\mathrm{1D}} + \frac{i}{k}\right]\,, \label{eq:T_1D_s} \\
a_{s,\mathrm{1D}} =& -\frac{a_{\perp}}{2}\left[\zeta\left(-\frac{1}{2}\right) + \zeta\left(\frac{1}{2}\right) \right. \nonumber \\
&\left. + \frac{a_{\perp}^3}{6} \left(\frac{1}{V_{p}} + \frac{4}{a_{\perp}^2 R_{p}}\right)\right] \,,\label{eq:1d_as} \\
r_{s,\mathrm{1D}} =& \frac{a_{\perp}^3}{16} \left[\zeta\left(\frac{1}{2} \right) + \zeta\left(\frac{3}{2} \right)  + \frac{4a_{\perp}}{3 R_{p}}\right]\,,
\label{eq:1D_rs}
\end{align}

\noindent which is equivalent to a true 1D calculation with even-wave interactions. The only difference between Eq.~(\ref{eq:T_1D_s}) and a true 1D calculation is that the matrix element for the low-dimensional $s$-wave scattering has an extra factor of two from the indistinguishability of the spin-polarized fermions. The parameters $a_{s,\mathrm{1D}}$ and $r_{s,\mathrm{1D}}$ in Eqs.~(\ref{eq:1d_as}-\ref{eq:1D_rs}) are the 1D $s$-wave scattering length and effective range, respectively. To the best of our knowledge, Eqs.~(\ref{eq:1d_as}-\ref{eq:1D_rs}) have only been reported in Ref.~\cite{Jackson23}.

\section{p-wave Scattering in q2D}
\label{sec:q2D}

Next, let us consider $p$-wave scattering in a q2D geometry. In this case, the eigenstates and energies for the Schr\"{o}dinger equation in the relative coordinates are:
\begin{align}
\psi_{k,n,m}({\bf r}_{\parallel},r_{\perp}) &= \frac{e^{i {\bf k \cdot r_{\parallel}}}}{\pi^{1/4}\sqrt{a_{z}}}\frac{e^{-\frac{r_{\perp}^2}{2a_z^2}} H_{2n+|m|}\left(\frac{r_{\perp}}{a_{\perp}}\right)}{\sqrt{2^{2n+|m|} (2n+|m|)!}}\,, \label{eq:q2d_basis} \\
E_{n,m}(k) &= \frac{k^2}{2\mu} + \left(2n + m+ \frac{1}{2}\right) \omega_{\perp}\,.
\label{eq:q2d_energy}
\end{align}
In Eqs.~(\ref{eq:q2d_basis}-\ref{eq:q2d_energy}), ${\bf k}$ is the momentum in the axial direction, while $n$ is the harmonic oscillator quantum number in the transverse directions. We have also defined the quantum number $m$, which controls the parity of the transverse wavefunction $P = (-1)^m$ and which takes on values of $m = 0,1$.

The matrix elements of the interaction are:
\begin{align}
    \langle &k n m |v | k' n' m\rangle = \nonumber \\ 
    &\frac{2g}{\sqrt{\pi} a_{\perp}}\left(f_s(n)f_s(n')  {\bf k \cdot k'} \delta_{m,0} + \frac{2}{a_{\perp}^2} f_p(n) f_p(n') \delta_{m,1}\right)\,,
\end{align}
where 
\begin{align}
f_s(n) &= \left(-\frac{1}{2}\right)^n \frac{\sqrt{(2n)!}}{n!}\,, \\
f_p(n) &= \left(-\frac{1}{2}\right)^n \frac{\sqrt{(2n+1)!}}{n!}\,.
\label{eq:q2D_pwave}
\end{align}

As in the q1D case, when $m=0$, the interaction matrix element is 1D $p$-wave in character, while for $m=1$ it is $s$-wave in character. The parity in the transverse direction acts equivalently to the angular momentum in the q1D geometry. For low-energies, $m=0$ occurs when both atoms are in the ground band, while $m=1$ is possible only when one of the atoms is in a higher transverse state.

The calculation for the q2D T-matrices is similar to the case of q1D scattering and provides the following expressions for $m = 0$:
\begin{align}
    \langle k n 0 |&T |k' n' 0 \rangle = 2 f_p(n)f_p(n') {\bf k \cdot k'} T_{p,\mathrm{2D}}(E)\,, \nonumber \\
    T_{p,\mathrm{2D}}^{-1}&=\frac{2\mu }{8\pi} \left[\frac{2\sqrt{\pi} a_{\perp}}{3} \left(\frac{1}{V_p} + \frac{2\mu E}{R_p}\right) + \frac{4}{a_{\perp}^2}w_0(\mathcal{E}_0)\right]\,,
    \label{eq:T0_2D}
\end{align}
and similarly for $m=1$:
\begin{align}
    \langle k n 1 | &T |k' n' 1 \rangle = 2 f_s(n)f_s(n') T_{s,\mathrm{2D}}(E)\,, \nonumber \\
    T_{s,\mathrm{2D}}^{-1}&=\frac{2\mu }{4\pi}\left[\frac{\sqrt{\pi} a_{\perp}^3}{6} \left(\frac{1}{V_p} + \frac{2\mu E}{R_p}\right) - w_1(\mathcal{E}_1)\right]\,.
    \label{eq:T1_2D}
\end{align}
We also define $\mathcal{E}_m = (E-(m+1/2)\omega_{\perp})/(2\omega_{\perp})$ for the q2D case, and the functions $w_{0,1}(\mathcal{E}_0)$ are defined as:
\begin{align}
w_0(\mathcal{E}_0) &= \sum_{n=0}^{N} \frac{(2n-1)!!}{(2n)!!}(n-\mathcal{E}_0) \ln(n- \mathcal{E}_0 - i \delta) - \mbox{norm.} \label{eq:w_0}\,,\\\
w_1(\mathcal{E}_1) &= \sum_{n=0}^{N} \frac{(2n+1)!!}{(2n)!!}\ln(n- \mathcal{E}_0 - i \delta) - \mbox{norm.} \label{eq:w_1}\,,
\end{align}
where $N$ is a high-energy cutoff which we send to infinity, and ``norm.'' represents the normalization terms. The normalization of Eqs.~(\ref{eq:w_0}-\ref{eq:w_1}) are discussed further in Appendix \ref{app:LSE}.

In the low-energy limit, $\mathcal{E}_m \ll 1$, these two-expressions collapse onto the desired expressions for 2D $s$- and $p$-wave T-matrices. Namely for $m=0$ we find the T-matrix for 2D $p$-wave interactions:
\begin{equation}
T_{p,\mathrm{2D}}^{-1}(E) = \frac{2\mu }{8\pi} \left[\frac{1}{a_{p,\mathrm{2D}}} + 2\mu E \ln\left(\frac{1}{-2\mu E r_{p,\mathrm{2D}}^2-i \delta}\right) \right]\,, \label{eq:T_2D_p}
\end{equation}
with the 2D scattering area $a_{p,\mathrm{2D}}$ and effective range $r_{p,\mathrm{2D}}$ given by
\begin{align}
\frac{a_{\perp}^2}{a_{p,\mathrm{2D}}} &= \frac{2 \sqrt{\pi}	a_{\perp}^3}{3} \left(\frac{1}{V_p}+ \frac{1}{a_{\perp}^2 R_p} \right) - \alpha_a \label{eq:a_p_2D_def}\,, \\
\ln\left(\frac{a_{\perp}}{r_{p,\mathrm{2D}}}\right) &= \frac{ \sqrt{\pi} a_{\perp}}{3 R_P} + \alpha_r\label{eq:r_p_2D_def}\,,
\end{align}
where $\alpha_a \approx 0.2173$ and $\alpha_r \approx 0.4689$. Eqs.~(\ref{eq:T_2D_p}-\ref{eq:r_p_2D_def}) were derived previously in Ref.~\cite{Zhang17}. 

Similarly for $m=1$ we obtain the T-matrix for 2D $s$-wave interactions:
\begin{equation}
T_{s,\mathrm{2D}}^{-1} (E) = \frac{2\mu }{4\pi} \left[\ln\left(\frac{1}{-2\mu E a_{s,\mathrm{2D}}^2 - i \delta}\right) + 2\mu E r_{s,\mathrm{2D}}\right]\,, \label{eq:T_2D_s}
\end{equation}
\begin{align}
\ln\left(\frac{a_{\perp}^2}{a_{s,\mathrm{2D}}^2}\right) &= \frac{\sqrt{\pi}a_{\perp}^3}{6} \left(\frac{1}{V_p} + \frac{3}{a_{\perp}^2 R_p}\right) + \ln \left(\frac{2B}{\pi}\right)\,, \label{eq:a_s_2D_def} \\
\frac{r_{s,\mathrm{2D}}}{a_{\perp}^2} &= {\frac{\sqrt{\pi}a_{\perp}}{6 R_P} + \frac{ \gamma}{4}}\,,
\label{eq:r_s_2D_def}
\end{align}
with $B \approx0.85$ and $\gamma \approx 1.41$.

\section{Two-Body Bound States}
\label{sec:bound states}

Since the T-matrices discussed in Secs. \ref{sec:q1D}-\ref{sec:q2D} are exact solutions to the two-body scattering, we can calculate exactly the two-body bound state and examine how it evolves from the low-dimensional to the 3D limits. This can be readily done by examining the pole of the two-body T-matrices in q1D and q2D.

First consider the bound-state equation in 3D. In practice there are three two-body bound states corresponding to corresponding to $m = -1,0,1$, two of which $m = \pm 1$ are degenerate in experiments \cite{Jackson23} due to the anisotropy of the dipole-dipole interaction. For our purposes we assume the interaction is isotropic so the three bound states are degenerate. In this case, the three-dimensional binding energy $E=-E_{B,\mathrm{3D}}$ for the two-body bound state is the solution of
\begin{equation}
    E_{B,\mathrm{3D}} = \frac{R_p}{2\mu V_p} + \frac{R_p}{2\mu} \left(2\mu E_{B,\mathrm{3D}}\right)^{3/2} 
    \label{eq:3D_BS}
\end{equation}
for $V_p > 0$. In the shallow-dimer limit ($R_p^{3/2} |V|^{-1/2} \ll 1$), the leading solution is \mbox{$E_{B,\mathrm{3D}}^0 = R_p/(2\mu V_p)$}. Also, for $V_p < 0$, a positive-energy scattering resonance exists at exactly $-E_{B,\mathrm{3D}}^0$.  

\subsection{q1D Bound State}

For the case of two-body bound states in the presence of harmonic confinement, it is convenient to introduce the bound state energies as: 
\begin{equation}
E_B^{(m)} = (|m|+1) \omega_{\perp} - \frac{\kappa_{m}^2}{2\mu}\,,
\end{equation}
where $\kappa_m^2/2\mu$ is the associated binding energy. The equations determining the bound state energies can then be written as:
\begin{equation}
-E_{B,\mathrm{3D}}^0 =E_B^{(0)}
+6 \frac{R_{p}}{a_{\perp}} \omega_{\perp} \zeta\left(-\frac{1}{2}, \left(\frac{\kappa_0 a_{\perp}}{2}\right)^2 \right) \,,
\label{eq:bs_odd}
\end{equation}
for odd-wave interactions and:
\begin{align}
-E_{B,\mathrm{3D}}^0 &= E_B^{(1)}+3 \frac{R_{p}}{a_{\perp}} \omega_{\perp}\left[\zeta\left(-\frac{1}{2}, \left(\frac{\kappa_{1} a_{\perp}}{2}\right)^2 \right) \right. \nonumber\\
&\left. + \left(1 - \left(\frac{\kappa_1 a_{\perp}}{2}\right)^2\right) \zeta \left(\frac{1}{2}, \left(\frac{\kappa_1 a_{\perp}}{2}\right)^2 \right)\right]\,,
\label{eq:bs_even}
\end{align}
for even-wave interactions. The right hand side of Eqs.~(\ref{eq:bs_odd}-\ref{eq:bs_even}) increases monotonically as a function of the binding energy, $\kappa_{0,1}^2/2\mu$. Hence there is always a bound state for $V_{p}<0$, which corresponds to a three-dimensional quasi-bound state. This is especially evident when one considers the limit: $\kappa_{0,1}^2/2\mu\gg1$. In this case one can show that the Hurwitz zeta functions for both p-wave and s-wave scattering yield a term that scales as $\kappa_{0,1}^{3}$. This is exactly the last term in Eq.~\eqref{eq:3D_BS} (as $\kappa^3 \sim E_B^{3/2}$), thus recovering the 3D equation for the two-body bound states.

\begin{figure}[tb]
\includegraphics[scale=0.5]{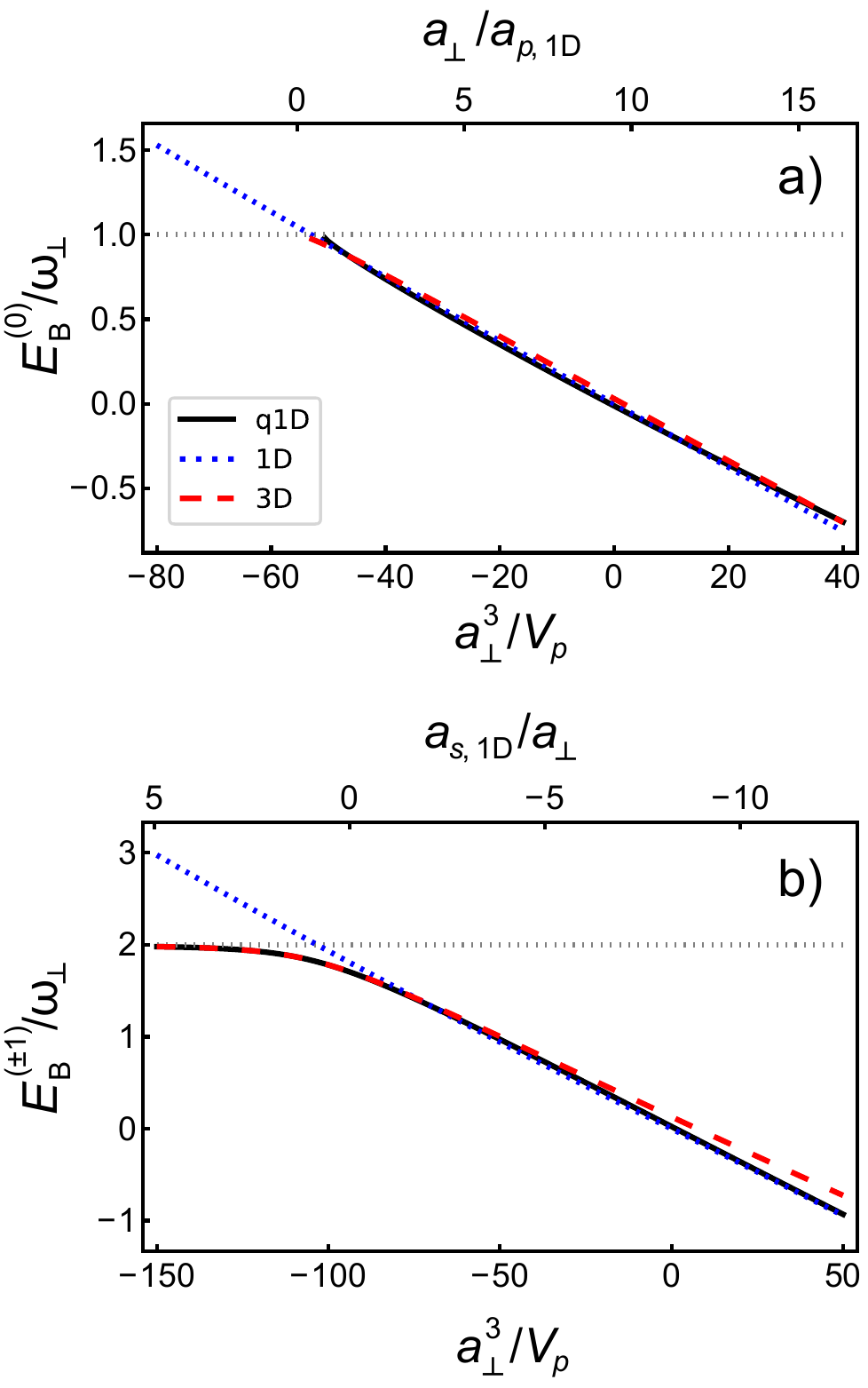}
\caption{$p$- (a) and $s$-wave (b) bound state energies plotted as function of both the scattering volume, and the respective 1D scattering parameters [Eqs.~\eqref{eq:1d_ap} and \eqref{eq:1d_as}]. The q1D results are obtained by solving Eqs.~(\ref{eq:bs_odd}-\ref{eq:bs_odd}) for the bound state energy. The 1D and 3D limits are shown by the red dashed and blue dotted lines respectively. The low-dimensional threshold is also shown as a black dotted line.}
\label{fig:q1D_BE}
\end{figure}

On the other hand, in the limit of shallow bound states, i.e. when $\kappa_{0,1}^2/2\mu \ll 2\omega_{\perp}$, Eqs.~(\ref{eq:bs_odd}-\ref{eq:bs_even}) reduce to the expected results for a 1D system with $p$- and $s$-wave interactions. In the $p$-wave case, we find that the equation determining $\kappa_0$ and the $p$-wave binding energy is:
\begin{equation}
    \frac{1}{a_{p,\mathrm{1D}}}- \kappa_0 - \kappa_0^2 r_{p,\mathrm{1D}}=0\,.
\end{equation}
Similarly, for $s$-wave case, we have
\begin{equation}
    -a_{s,\mathrm{1D}} + \frac{1}{\kappa_1} - \kappa_1^2 r_{s,\mathrm{1D}}=0\,.
\end{equation}

Thus Eqs.~(\ref{eq:bs_odd}-\ref{eq:bs_even}) describe the behavior of the two-body bound state over the whole dimensional crossover. This is exemplified in Fig.~\ref{fig:q1D_BE} which presents the results for the full dimensional crossover for the binding energy, i.e. Eqs.~(\ref{eq:bs_odd}-\ref{eq:bs_even}), alongside the limiting low-dimensional and 3D behaviours. As one can see, the full solution smoothly interpolates between both limiting behaviors as a function of the 3D scattering volume or the respective 1D scattering parameter.

\subsection{q2D Bound State}
In q2D, the two-body bound state energies are defined as:
\begin{equation}
    E_B^{(m)} = \left(m+\frac{1}{2}\right)\omega_{\perp} -\frac{\kappa_m^2}{2\mu}
\end{equation}

it is beneficial to solve the T-matrix in a different approach. Eqs.~(\ref{eq:T0_2D}-\ref{eq:T1_2D}) are the result of integrating over  all the virtual states in the unconfined two-dimensional plane, describing the scattering in q2D as a discrete set of contributions which come from the discrete nature of the states in the confined direction. 

To study the behavior of the bound state across the dimensional crossover, however, it is more practical to do the reverse, and obtain an expression for the q2D T-matrices which have integrated out the discrete transverse excitations, and leaving an expression in terms of the virtual q2D plane wave modes \cite{LevinsenParish:2015}. This is an alternative but identical representation of the q2D T-matrix. The details are discussed in Appendix \ref{app:parish}. Here we present the equations for the binding energies:
\begin{widetext}
\begin{equation} \label{eq:2DpwaveDimer} 
\begin{aligned}
\frac{-E_{B,3D}^0}{\omega_{\perp}} = \frac{E_B^{(0)}}{\omega_{\perp}} + \frac{3}{4\pi^{1/2}} \frac{R_p}{a_\perp} \int_0^\infty \!\! du  & \left[ \frac{u e^{-\mathcal{E}_0 u} }{\lambda u + 4 \lambda^2} \frac{\frac{1}{2}(1-\lambda)^2 e^{-u} + \mathcal{E}_0 ((1+\lambda)^2 - e^{-u}(1-\lambda)^2)}{((1+\lambda)^2 - e^{-u}(1-\lambda)^2)^{3/2}} \right. \\
& - \left. \frac{4}{(u+ 4 \lambda)^{5/2}} - \frac{8}{3} \frac{1/4-\mathcal{E}_0}{(u+4 \lambda)^{3/2}} \right]\,,
\end{aligned} 
\end{equation}
\\
\begin{equation} \label{eq:2DswaveDimer}
\frac{-E_{B,3D}^0}{\omega_{\perp}} = \frac{E_B^{(1)}}{\omega_{\perp}} + \frac{3}{ \pi^{1/2}} \frac{R_p}{a_\perp} \int_0^\infty \!\! du \left[ \frac{e^{-\mathcal{E}_1 u} }{u + 4 \lambda} \frac{1}{[(1+\lambda)^2 - e^{-u}(1-\lambda)^2]^{3/2}} - \frac{1}{(u+ 4 \lambda)^{5/2}} - \frac{2}{3} \frac{3/4 -\mathcal{E}_1}{(u+4 \lambda)^{3/2}} \right]\,,
\end{equation}
\end{widetext}
where \mbox{$\mathcal{E}_{m} = \kappa_m^2/(4\mu \omega_{\perp})$} is the dimensionless binding energies. We also define $\lambda=2(\Lambda a_\perp)^{-2}$ with $\Lambda$ as a cutoff parameter which in our model is given by $R_p^{-1}$. The left hand side of each equation is the leading low-energy binding energy in the 3D limit, while the first two terms on the right hand side describe the quasi-low-dimensional bound state relative to the zero-point energy. The integral on the left hand side contains two pieces: first is the contribution due to virtual excitations for a quasi-low-dimensional systems, and the second are terms represent the renormalization of the 3D interaction.

In the limit of small binding energies $\mathcal{E}_{0,1} \ll 1$, we expect the dominant contribution to the integrals in Eqs.~(\ref{eq:2DpwaveDimer}-\ref{eq:2DswaveDimer}) come from the virtual excitation term. An analysis of the integral in this q2D limit, described in Appendix \ref{app:parish}, yields the expected low-dimensional expressions for the binding energies:
\begin{align}
\kappa_0^2 r_{p,\mathrm{2D}}^2 \ln\left(\kappa_0^2 r_{p,\mathrm{2D}}^2\right) &= - \frac{r_{p,\mathrm{2D}}^2}{a_{p,\mathrm{2D}}}\,, & m=0\,,\\
\frac{1}{\kappa_1^2 a_{s,\mathrm{2D}}^2}\ln\left(\frac{1}{\kappa_1^2 a_{s,\mathrm{2D}}^2}\right) &=  \frac{r_{s,\mathrm{2D}}}{a_{s,\mathrm{2D}}^2}\,, & m=1\,.
\end{align}
Above we have used the definitions for the two-dimensional scattering parameters, Eqs.~(\ref{eq:a_p_2D_def}-\ref{eq:r_p_2D_def}) and Eqs~ (\ref{eq:a_s_2D_def}-\ref{eq:r_s_2D_def}). 
These equations can be solved using Lambert's W functions:
\begin{align}
    \kappa_0^2 &= -\frac{1}{a_{p,\mathrm{2D}}} W_0^{-1}\left(-\frac{r_{p,\mathrm{2D}}^2}{a_{p,\mathrm{2D}}}\right), \\
    \kappa_1^2 &= \frac{1}{r_{s,\mathrm{2D}}} W_0\left(\frac{r_{s,\mathrm{2D}}}{a^2_{s,\mathrm{2D}}}\right).
\end{align}

\begin{figure}[tb]
\includegraphics[scale=0.5]{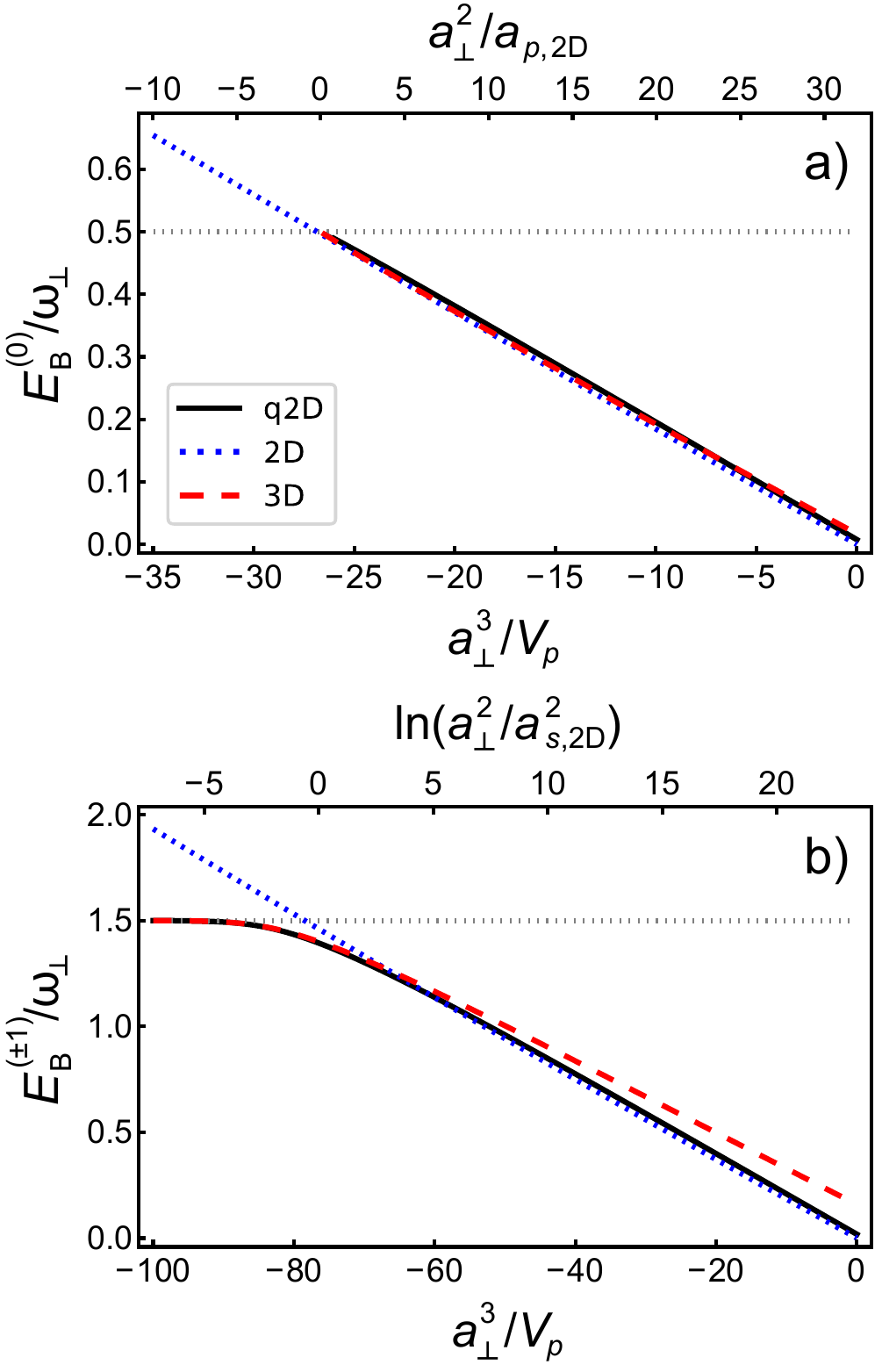}
\caption{$p$- (a) and $s$-wave (b) bound state energies plotted as function of both the scattering volume, and the respective 2D scattering parameters [Eqs.~(\ref{eq:a_p_2D_def} and \ref{eq:a_s_2D_def})]. The 2D and 3D limits are shown by the red dashed and blue dotted lines respectively. Again we find the q2D bound state formulae Eqs.~(\ref{eq:2DpwaveDimer}-\ref{eq:2DswaveDimer}) interpolates between the low- and high-dimensional limits.}
\label{fig:q2D_BE}
\end{figure}

In the opposite limit of deep dimers $\mathcal{E}_{s,p} \gg 1$, we expect that both the virtual excitation term and renormalization terms are important. As shown in Appendix \ref{app:parish}, a similar analysis in this limit gives results consistent with the 3D prediction, Eq.~\eqref{eq:3D_BS}.

We present the numerical solution of Eqs.~(\ref{eq:2DpwaveDimer}-\ref{eq:2DswaveDimer}) in Fig.~\ref{fig:q2D_BE}, alongside the predictions for the 3D and low-energy 2D theories. In these solutions the effective range was chosen to match the parameterization of the commonly used p-wave resonance in $^{40}$K \cite{Ahmed-Braun21}. We find that the both the $p$-wave and $s$-wave q2D bound states smoothly cross over from the low- to high-dimensional results with the intermediate condition being $\mathcal{E}_{s,p} \sim 1$, or $\kappa_{s,p}a_{\perp} \sim 1$. Ref.~\cite{exp_companion} presents experimental evidence for both the $p$-wave and $s$-wave dimers in q2D $^{40}$K, whose energies are as predicted by Eqs.~(\ref{eq:2DpwaveDimer}-\ref{eq:2DswaveDimer}). 

\section{RF Spectroscopy of Spin-Polarized Fermi Gases}
\label{sec:RF}

We now apply the results for the two-body scattering to the calculation of the RF spectroscopy transfer rate in the high frequency limit. A RF pulse induces a hyperfine-Zeeman spin flip from state $|a \rangle$ to state $|b \rangle$ at a fixed detuning, $\omega$. The transfer rate for this process is defined as \cite{Braaten10, Son10}:
\begin{align}
&\Gamma(\omega) = \frac{\Omega^2}{4}{\rm Im} \ i \int_{0}^{\infty} dt \int d^3 {\bf R} \int d^3 {\bf r} \ e^{i(\omega+i \delta)t}  \nonumber \\
&\times\left\langle T_t \ \psi_a^{\dagger}\psi_b\left({\bf R} + \frac{1}{2}{\bf r}, t\right)\psi_b^{\dagger}\psi_a\left({\bf R} - \frac{1}{2}{\bf r}, 0\right) \right\rangle\,, \label{eq:RF}
\end{align}
where $\Omega$ is the Rabi frequency, and $\psi_a^{\dagger}({\bf r})$ creates one fermions in hyperfine-Zeeman state $a$.

\subsection{Two-Body Solution}
For large detuning, the RF pulse probes the short-distance physics of the many-body wave function which is dominated by two-body correlations. Thus we proceed to evaluate the RF transfer rate at the two-body level. For both q1D and q2D geometries, the RF transfer rate has a contribution from each value of $m$:
\begin{equation}
\Gamma(\omega) = \sum_m \Gamma^{(m)}(\omega)\,,
\end{equation}
which can be evaluated independently. The corresponding RF transfer rate  at the two-body level is determined by the diagrams shown in Fig.~\ref{fig:OPE}. The two-body calculation is presented in  Appendix \ref{app:ope_appendix}. Here we present the results for q1D:
\begin{align}
\Gamma^{(0)}(\omega) &\approx  \Omega^2\frac{A_0(E_F,T)}{8\pi} \frac{(2\omega_{\perp})^{3/2}}{\omega^2} \nonumber \\
&\times\sum_{n=0}^{\infty} \sqrt{\frac{\omega}{2\omega_{\perp}}-n}\ \theta\left(\frac{\omega}{2\omega_{\perp}}-n\right) 
\label{eq:RF_2B_odd_1D}\,, \\
\Gamma^{(\pm 1)}(\omega) &\approx \Omega^2\frac{A_{\pm 1}(E_F,T)}{16\pi} \frac{(2\omega_{\perp})^{3/2}}{\omega^2} \nonumber \\
&\times\sum_{n=0}^{\infty} (n+1)\left(\frac{\omega}{2\omega_{\perp}}-n\right)^{-1/2}\ \theta\left(\frac{\omega}{2\omega_{\perp}}-n\right)\,,
\label{eq:RF_2B_even_1D}
\end{align}
where the constants $A_m(E_F,T)$ are the sole fitting parameters which describe the many body physics, i.e. dependent on the Fermi energy, $E_F$, and the temperature, $T$. We have also assumed that all the atoms scatter in the lowest $p$-wave and $s$-wave scattering continua. Similarly in q2D one finds:
\begin{align}
    \Gamma^{(0)}(\omega) &\approx \Omega^2\frac{A_0(E_F,T)}{16\sqrt{\pi}} \frac{(2\omega_{\perp})^{3/2}}{\omega^2} \nonumber \\
    &\sum_{n=0}^{\infty} \frac{(2n-1)!!}{(2n)!!} \left(\frac{\omega}{2 \omega_{\perp}}-n\right)\theta\left(\frac{\omega}{2\omega_{\perp}}-n\right)\,,
    \label{eq:RF_2B_odd_2D} \\
    \Gamma^{(1)}(\omega) &\approx \Omega^2 \frac{A_1(E_F,T)}{16\sqrt{\pi}} \frac{(2\omega_{\perp})^{3/2}}{\omega^2} \nonumber \\
    &\sum_{n=0}^{\infty} \frac{(2n+1)!!}{(2n)!!} \theta\left(\frac{\omega}{2\omega_{\perp}}-n\right)\,.
    \label{eq:RF_2B_even_2D}
\end{align}
We note that similar results were obtained for the case of q2D Fermi gases with $s$-wave interactions \cite{Baur12}.

\begin{figure}[tb]
\includegraphics[scale=0.6]{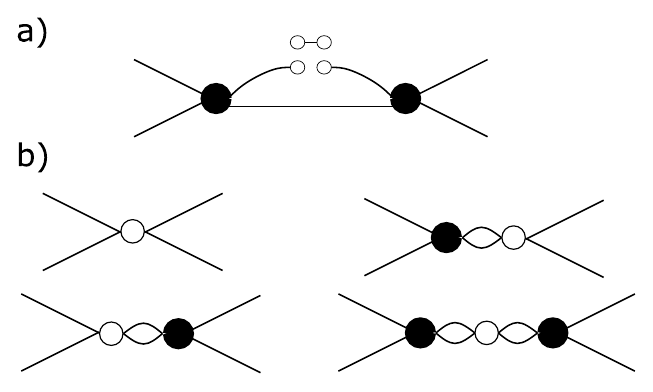}
\caption{Feynman diagrams contributing to a) the two-body solution to the RF transfer rate and b) the right hand side of the OPE, see Eq.~(\ref{eq:ope_def}). The black circles represent the two-body T-matrices in the presence of harmonic confinement. In a) the open circles is the operator insertion in Eq.~(\ref{eq:RF}), while in b) it represents the insertion of a contact operator, Eq.~(\ref{eq:3D_contacts}).}
\label{fig:OPE}
\end{figure}

\begin{figure}[tb]
\includegraphics[scale=0.55]{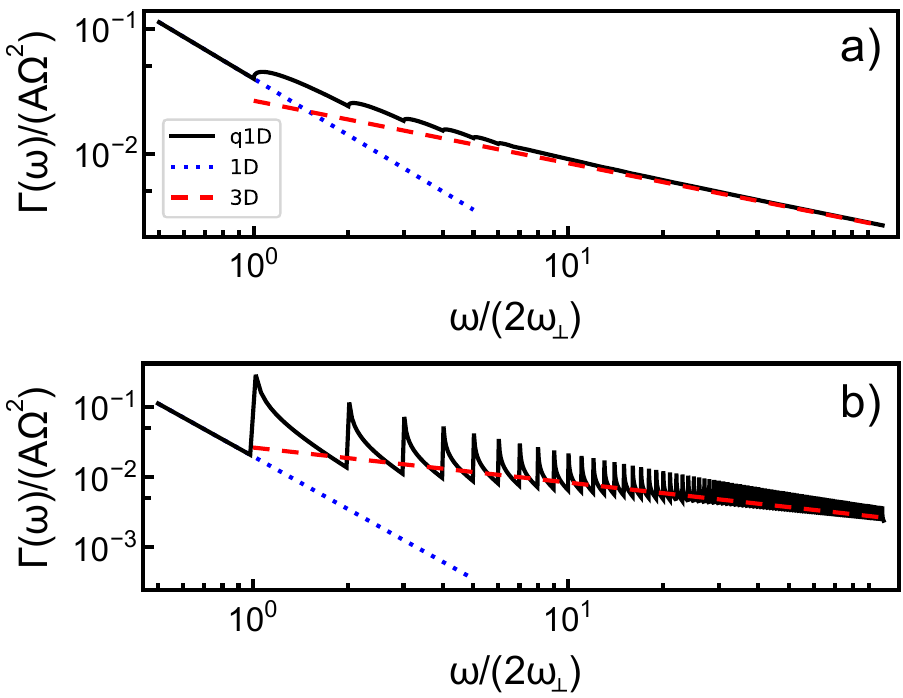}
\caption{$p$- (a) and $s$-wave (b) contributions to the q1D RF spectroscopy rate as a function of detuning, $\omega$, Eqs.~(\ref{eq:RF_2B_odd_1D}-\ref{eq:RF_2B_even_1D}), on a log-log-scale. The 3D and 1D limits are shown by the red dashed and blue dotted lines respectively. The kinks (divergences) for odd-wave (even-wave) scattering occur whenever a new $p$-wave ($s$-wave) scattering continuum becomes energetically available. These are universal features that become smeared out in the 3D limit.}
\label{fig:q1D_RF}
\end{figure}

\begin{figure}[tb]
\includegraphics[scale=0.55]{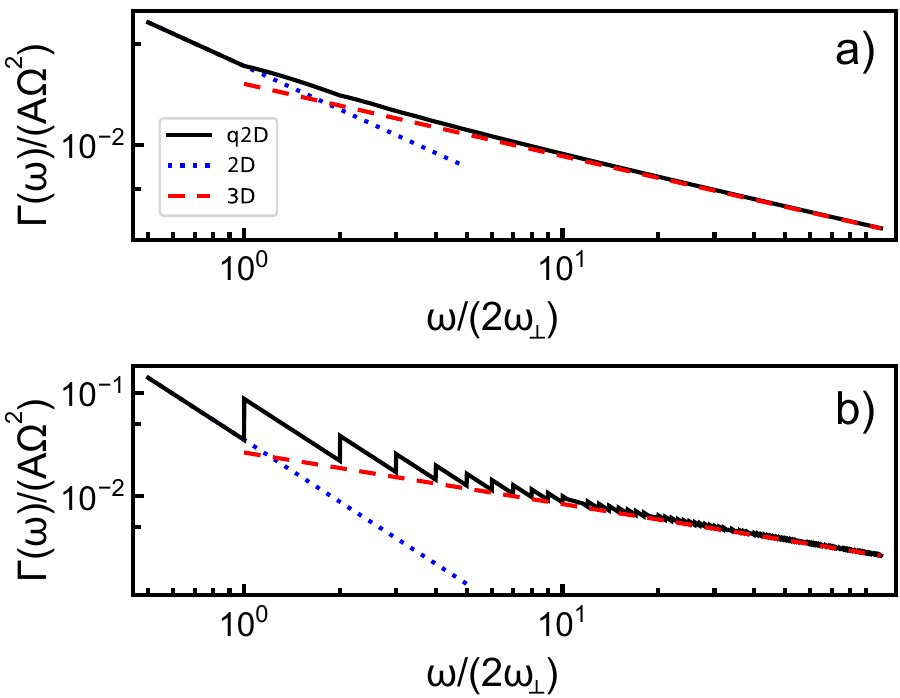}
\caption{$p$- (a) and $s$-wave (b) contributions to the q2D RF spectroscopy rate as a function of detuning, $\omega$, Eqs.~(\ref{eq:RF_2B_odd_2D}-\ref{eq:RF_2B_even_2D}), on a log-log-scale. The 3D and 2D limits are shown by the red dashed and blue dotted lines respectively. The kinks (steps) for $p$-wave ($s$-wave) are related to new harmonic states becoming energetically available. }
\label{fig:q2D_RF}
\end{figure}

Eqs.~(\ref{eq:RF_2B_odd_1D}-\ref{eq:RF_2B_even_2D}) are shown in Figs.~(\ref{fig:q1D_RF}-\ref{fig:q2D_RF}) for a wide range of detunings. 
In the limit $\omega \ll 2\omega_{\perp}$, only the initial low-dimensional scattering continuum contributes to the RF transfer rate. In this limit one finds that the RF transfer rate is consistent with the true low dimensional calculation, as shown in Appendix \ref{appendix:review}, as shown by the blue lines in Figs.~(\ref{fig:q1D_RF}-\ref{fig:q2D_RF}). This scaling appears to be robust all the way up to $\omega = 2\omega_{\perp}$, when a new low-dimensional scattering continuum becomes energetically available. The availability of a new scattering continuum leads to a non-monotonic behavior in the RF transfer rate, as is shown in Fig.~\ref{fig:q1D_RF} for q1D, and Fig.~\ref{fig:q2D_RF} for q2D. The scaling behavior of this discontinuity only depends on the low-energy behavior of the $s$- and $p$-wave  scattering, and the geometry of the system, as can be illustrated by Fermi's golden rule. 

Consider an RF spin flip from an initial state $|i \rangle = |k, n_i,m\rangle$ to a final state, $|f\rangle = |k,n_f,m\rangle$. The matrix element for the RF spin flip operator, $V_{RF}$ is related to the Franck-Condon factor, i.e. the overlap between the initial and final scattering states: $|\langle f | V_{RF} | i \rangle|^2 \propto k^{2(1-|m|)}$. Given that the axial density of states scales like $1/k$ for 1D and $k^0$ for 2D, Fermi's golden rule states that the RF transfer rate is: $\propto k$ for $p$-wave scattering and $\propto k^{-1}$ for $s$-wave scattering in q1D. Similarly the RF transfer rate is proportional to $k^2$ for $p$-wave scattering and $k^0$ for $s$-wave scattering in q2D. In the RF process, the excess kinetic energy is related to $k^2/2\mu =\delta \omega = \omega - 2n_f \omega_{\perp}$. The universal discontinuity in the RF transfer rate then has the the form $\sqrt{\delta \omega}$ for the $p$-wave sector and $1/\sqrt{\delta \omega}$ for $s$-wave interactions in q1D. In q2D these discontinuities are of the following form: $\delta \omega$ for $p$-wave scattering and $(\delta \omega)^0$ for $s$-wave scattering. This discontinuity is directly captured in the two-body solutions, and can be clearly seen in Figs.~(\ref{fig:q1D_RF}-\ref{fig:q2D_RF}).

In the 3D limit many effective low-dimensional scattering continua become open. In this limit we can replace the sum over transverse states with a continuous integral. In both q1D and q2D the RF transfer rates then converge on the three-dimensional result:
\begin{align}
\Gamma^{(0)}(\omega) &= \Omega^2 A_0(E_F,T)\frac{1}{12\pi \sqrt{\omega}}\,,  \\
\Gamma^{(\pm 1)}(\omega) &= \Omega^2 A_{\pm1}(E_F,T) \frac{1}{12\pi \sqrt{\omega}}\,.
\label{eq:3D_Limit}
\end{align}
Eq.~(\ref{eq:3D_Limit}) is consistent with the leading frequency behavior one would acquire from a 3D calculation with $p$-wave interactions, as detailed in Appendix \ref{appendix:review}, and is shown in Fig.~\ref{fig:q1D_RF} by the dashed red dashed line.

\subsection{Operator Product Expansion}

Up to this point we have only considered two-body correlation effects and their role on the RF transfer rate. To make this argument more rigorous, we can use the operator product expansion (OPE) to obtain the leading high-frequency behavior and the sub-leading tail. In the OPE we expand the time-ordered correlation function in a series of local operators. The local operators carry all the information of the many-body sate, while the series coefficients contain the information on the detuning $\omega$ \cite{Braaten10,Son10}: 
\begin{equation}
\Gamma(\omega) = \sum_a W_a(\omega) \langle O_a \rangle\,,
\label{eq:ope_def}
\end{equation}
where $\langle O_a\rangle$ is the thermodynamic average of a local operator, $O_a$, and $W_a(\omega)$ is the corresponding Wilson coefficient. This process is most easily done in the low-dimensional and 3D limits. 

The operators relevant to our discussion are the contact operators in  D dimensions, which describe how the energy changes with respect to the low-energy scattering parameters. The $d$-dimensional $s$- and $p$-wave contact operators are summarized in Appendix \ref{appendix:review}. In 3D, there are two contact operators:
\begin{align}
C_{V_{p}} &= -\frac{\partial H}{\partial V_{p}^{-1}}\,, & C_{R_{p}} &= -\frac{\partial H}{\partial R_{p}^{-1}}\,.
\label{eq:3D_contacts}
\end{align}
As discussed in Appendix \ref{app:ope_appendix}, one can evaluate the expectation values of the contact operators and compare to the two-body solutions of the RF-transfer rate, Eq.~(\ref{eq:RF_2B_odd_1D}-\ref{eq:RF_2B_even_2D}) to extract the Wilson coefficients. In the 3D limit, it is straight forward to show that both for q1D and q2D:
\begin{equation}
\left.\Gamma(\omega)\right|_{3D} \approx \frac{2\Omega^2}{\sqrt{\omega}}\left[\langle C_{V_{p}} \rangle + \frac{3}{2\omega} \langle C_{R_{p}}\rangle\right]\,,
\label{eq:RF_3D}
\end{equation}
which is the correct 3D result. In the low-dimensional limit, one obtains:
\begin{align}
\left.\Gamma^{(0)}_{p}(\omega)\right|_\mathrm{1D} &\approx \frac{\Omega^2}{2\omega^{3/2}} \frac{6}{a_{\perp}^2} \left[ \langle C_{V_p}^{(0)} \rangle \right. \nonumber \\
&\left. + \frac{1}{2\omega} \left( \langle C_{R_p}^{(0)} \rangle - \frac{2}{a_{\perp}^2} \langle C_{V_p}^{(0)} \rangle\right) \right]\,, \label{eq:Gamma_q1d_p_1d} \\
\left.\Gamma^{(\pm 1)}(\omega)\right|_\mathrm{1D} &\approx \frac{\Omega^2}{4\omega^{5/2}} \frac{24}{a_{\perp}^4} \langle C_{V_p}^{(\pm 1)} \rangle\,, \label{eq:RF_1D_3D} 
\end{align}
for q1D systems, and similarly for q2D systems:
\begin{align}
\left.\Gamma^{(0)}(\omega)\right|_\mathrm{2D} &= \frac{\pi\Omega^2}{2\omega} \frac{3}{2 \sqrt{\pi} a_{\perp}} 
\left[ \langle C_{V_p}^{(0)} \rangle \right. \nonumber \\
& \left.+ \frac{1}{\omega} \left( \langle C_{R_p}^{(0)} \rangle - \frac{1}{a_{\perp}^2} \langle C_{V_p}^{(0)} \rangle\right) \right]\,, \label{eq:Gamma_q2d_p_2d} \\
\left. \Gamma^{(1)}(\omega)\right|_\mathrm{2D} &= \frac{\pi\Omega^2}{4\omega^2} \frac{12}{\sqrt{\pi}a_{\perp}^3} \langle C_{V_p}^{(1)} \rangle\,. \label{eq:RF_2D_3D} 
\end{align}

Eqs.~(\ref{eq:RF_1D_3D}-\ref{eq:RF_2D_3D}) describe the low-dimensional RF transfer rate in terms of the 3D contacts with definite angular momentum (parity) $m$: $C_{V_p}^{(m)}$ and $C_{R_p}^{(m)}$. The 3D contacts can then be connected to the low-dimensional ones using the definitions in Eqs.~(\ref{eq:1d_ap}-\ref{eq:1D_rs}) for the q1D scattering parameters and Eqs.~(\ref{eq:a_p_2D_def}-\ref{eq:r_s_2D_def}) for the q2D scattering parameters. The definitions for the low-dimensional contacts are presented in Appendix \ref{appendix:review}. Here we simply quote the results for q1D systems:
\begin{align}
C_{V_{p}}^{(0)} &= \frac{a_{\perp}^2}{6} C_{a_{p,\mathrm{1D}}} \,,\label{eq:contact_relations_q1D}\\
C_{R_p}^{(0)} &= \frac{1}{3} C_{a_{p,\mathrm{1D}}} + \frac{a_{\perp}^2}{6} C_{r_{p,\mathrm{1D}}} \,,\\
C_{V_{p}}^{(\pm 1)} &= \frac{a_{\perp}^4}{12} C_{a_{s,\mathrm{1D}}}\,, \\
C_{R_{p}}^{(\pm 1)} &= \frac{a_{\perp}^2}{3} C_{a_{s,\mathrm{1D}}} + \frac{a_{\perp}^4}{12}C_{r_{s,\mathrm{1D}}}\,,
\end{align}
and for q2D systems:
\begin{align}
C_{V_{p}}^{(0)} &= \frac{2 \sqrt{\pi}a_{\perp}}{3} C_{a_{p,\mathrm{2D}}}\,,\\
C_{R,p}^{(0)} &= \frac{2\sqrt{\pi}}{3a_{\perp}} C_{a_{p,\mathrm{2D}}} + \frac{2 \sqrt{\pi}a_{\perp}}{3} C_{r_{p,\mathrm{2D}}}\,,\\
C_{V_{p}}^{(\pm 1)} &= \frac{\sqrt{\pi}a_{\perp}^3}{6} C_{a_{s,\mathrm{2D}}}\,, \\
C_{R_{p}}^{(\pm 1)} &= \frac{\sqrt{\pi}a_{\perp}^2}{2} C_{a_{s,\mathrm{2D}}} + \frac{\sqrt{\pi}a_{\perp}^3}{6}C_{r_{s,\mathrm{2D}}}\,.
\label{eq:contact_relations_q2D}
\end{align}
Given the above equations, Eqns.~(\ref{eq:contact_relations_q1D}-\ref{eq:contact_relations_q2D}), one can show that the RF transfer rate collapses onto the desired result of a true low-dimensional system, see Appendix \ref{appendix:review}:
\begin{align}
    \Gamma_{p,\mathrm{1D}}(\omega) &= \frac{\Omega^2}{2\omega^{3/2}} \left[\langle C_{a_{p,\mathrm{1D}}} \rangle + \frac{1}{2\omega} \langle C_{r_{p,\mathrm{1D}}}\rangle \right]\,, \\
    \Gamma_{p,\mathrm{2D}}(\omega) &= \frac{\pi \Omega^2}{2\omega} \left[\langle C_{a_{p,\mathrm{2D}}} \rangle + \frac{1}{\omega} \langle C_{r_{p,\mathrm{2D}}}\rangle \right]\,, \\
    \Gamma_{s,\mathrm{1D}}(\omega) &= 2\frac{\Omega^2}{4\omega^{5/2}}\langle C_{a_{s,\mathrm{1D}}}\rangle\,, \\
    \Gamma_{s,\mathrm{2D}}(\omega) &= 2\frac{\pi\Omega^2}{4\omega^{2}} \langle C_{a_{s,\mathrm{2D}}}\rangle\,.
\end{align}
The only difference is that the RF transfer rate for the even-wave interactions is twice that of a true $s$-wave system, which comes from the indistinguishable nature of the spin-polarized fermions.

\section{Conclusions}
\label{sec:conc}

In this work we have analyzed the $p$-wave scattering in the presence of strong harmonic confinement potentials, typically used to create q1D or q2D systems. We found that there were two effective low-dimensional interactions, one that is $p$-wave and the other that is $s$-wave in character, and we solved the corresponding two-body problems. Both interactions are necessary in understanding the dimensional crossover in RF spectroscopy experiments. At the two-body level we found a complete description of the dimensional crossover, which is shown in Fig.~\ref{fig:q1D_RF} for the 1D-to-3D crossover, and in Fig.~\ref{fig:q2D_RF} for the 2D-to-3D crossover. To make these results more rigorous we used the OPE to relate the leading and sub-leading high-frequency power-law tails to the thermodynamic contacts in the low-dimensional and 3D limits.

These results can be tested experimentally. Already in Ref.~\cite{Jackson23} a clear signal was observed in the RF spectroscopy indicating the opening of a new $s$-wave scattering channel for a spin-polarized Fermi gas of $^{40}$K atoms in a q1D trapping geometry, matching Eq.~(\ref{eq:RF_2B_even_1D}). The experimental observation of q2D $s$-wave channel is reported in Ref.~\cite{exp_companion}. 
The non-monotonic behavior of the RF transfer function is quite general and depends only on the form of the interaction and the geometry of the system. Similar results were obtained for the RF spectroscopy of a q2D Fermi gas with $s$-wave interactions \cite{Fischer14}. 

Our analysis is quite general and can be also applied to a number of different scenarios. In particular understanding the dimensional crossover from spectroscopic probes should be possible in, for example, the structure factor at zero momentum, or the closed-channel fraction. This calculation can also be readily extended to study other dimensional crossovers with other higher partial wave interactions.

A particularly interesting result is the presence of different types of low-dimensional scattering, i.e. scattering with different parity or angular momentum, at higher energies. This opens up a rich new area of physics where atomic gases can occupy a finite number of transverse harmonic oscillator states. In such systems, the different low-dimensional interactions can compete, and can result in strong correlations, as observed in Ref.~\cite{Jackson23}.

\paragraph*{Acknowledgements}-- The authors would like to thank Kevin G. S. Xie for useful discussions. This work is supported by HK GRF Grants No. 17306024, CRF Grants No. C6009-20G and No. C7012-21G, and a RGC Fellowship Award No. HKU RFS2223-7S03. JM is supported in part by the Provincia di Autonoma di Trento.

\appendix

\numberwithin{equation}{section}
\renewcommand\theequation{\Alph{section}.\arabic{equation}}

\section{Solution of Lippmann-Schwinger Equation}
\label{app:LSE}

In this appendix, we solve the Lippmann-Schwinger equation in the presence of the external trapping potential that is described by the Hamiltonian given by Eq.~(\ref{eq:2B_Hamiltonian}). The inter-atomic potential $v(r)$ is given by the $p$-wave pseudo-potential, Eq.(\ref{eq:pseudopotential}). The derivation presented here follows closely that of Ref.~\cite{Moore04}.

We begin with the definition of the $T$-matrix in free space:
\begin{equation}
T_{3D}(E) = v + v G_{3D}^0(E) T_{3D}(E)\,,
\end{equation}
where $G_{3D}^0(E) =(E-H_0+i\delta)^{-1}$ is the bare propagator in three-spatial dimensions without confinement. Here $H_0$ is simply the kinetic energy. Given the low-energy T-matrix, one can solve for the bare interaction:
\begin{equation}
    v^{-1}=T_{3D}(E)^{-1}+G_{3D}^0\,.
\end{equation}
The bare interaction can similarly be solved in terms of the T-matrix in the presence of the confinement $T(E)$:
\begin{equation}
    v^{-1}=T(E)^{-1}+G^0(E)\,,
    \label{eq:v_trap}
\end{equation}
where now $G^0(E)$ is the bare propagator in the pressence of the confinement potential $H_t$: $G^0(E) = (E-H_0-H_t+i\delta)^{-1}$. Eliminating $v^{-1}$ which is identical in both free space and confined scattering, one obtains $T_{3D}(E)^{-1}+G_{3D}^0=T(E)^{-1}+G^0(E)$. Re-arranging, for $T(E)$ one can relate the scattering from some interaction $v$ inside a trapping potential to scattering to the same scattering in free space: 
\begin{equation}
T(E) = T_{3D}(E) + T_{3D}(E)\left(G^0(E) - G_{3D}^0(E) \right)T(E)\,.
\label{eq:Lupu_Sax}
\end{equation}
This result was first derived in Ref.~\cite{Moore04}.

For $p$-wave scattering it is possible to simplify Eq.~(\ref{eq:Lupu_Sax}) since both the 3D free space T-matrix and the T-matrix inside the harmonic trapping potential are separable:
\begin{equation}
    T_{(3D)}(E) = \overleftarrow{\nabla}  \cdot \tilde{T}_{(3D)}(E) \cdot \overrightarrow{\nabla}\,,
    \label{eq:T3D_position space}
\end{equation}
where the (3D) T-matrix has been written in position space, and the arrows denote whether the gradient is applied to the bra or the ket. Eq.~(\ref{eq:T3D_position space}) is consistent with the matrix elements of the pseudo-potential. Inspired by the separable form of the 3D T-matrix, we propose the following ansatz that solves Eq.~(\ref{eq:Lupu_Sax}):
\begin{align}
T(E) &= \overleftarrow{\nabla}  \cdot \tilde{T}(E) \cdot \overrightarrow{\nabla}\,, \nonumber \\
\tilde{T}(E)^{-1} &= \tilde{T}_{3D}^{-1}  \nonumber \\
&- \lim_{{\bf r}, {\bf r}'\to 0}\nabla_{\bf r} \cdot \nabla_{{\bf r}'} \left[G^0(E,{\bf r},{\bf r}')-G^0_{3D}(E,{\bf r},{\bf r'})\right]\,.
\label{eq:ansatz}
\end{align}
From here it is necessary to consider the specific structure of the trapped and free-space propagator. First consider q1D scattering. The non-interacting propagator for a q1D system:
\begin{align}\nonumber
G^0(E; r_{\parallel},{\bf r}_{\perp}; r_{\parallel}', {\bf r}_{\perp}') &=  
- \frac{2\mu a_{\perp}}{4}\sum_{m=-\infty}^{\infty}\sum_{n=0}^{\infty}\frac{\phi_{n,m}({\bf r}_{\perp}) \phi_{n,m}^*({\bf r}_{\perp}')}{\sqrt{n - \mathcal{E}_m- i \delta}}\\
&\times e^{-\frac{2}{a_{\perp}}\sqrt{n - \mathcal{E}_m- i \delta}|r_{\parallel}-r_{\parallel}'|}\,,
\label{eq:trapped_Propagator}
\end{align}
where $\mathcal{E}_m = (E-(|m|+1)\omega_{\perp})/(2\omega_{\perp})$, and $\phi_{n,m}({\bf r}_{\perp})$ is the wavefunction of a 2D harmonic oscillator, see Eq.~(\ref{eq:q1d_basis}). To make a direct connection to the trapped propagator, the free-space propagator must also be expanded in terms of modes with definite angular momentum, $m$:
\begin{align}\nonumber
&G_0(E; r_{\parallel},{\bf r}_{\perp}; r_{\parallel}', {\bf r}_{\perp}') =-\frac{2\mu}{4L} 
\sum_{m= -\infty}^{\infty} \sum_{n=0}^{\infty} \frac{k_{n,m}}{\sqrt{k_{n,m}^2 - 2\mu E- i \delta}}  \\\nonumber
&\times e^{-\sqrt{k_{n,m}^2 - 2\mu E - i \delta}|r_{\parallel}-r_{\parallel}'|}\\
&\times J_{|m|}(k_{n,m}r_{\perp})J_{|m|}(k_{n,m}r_{\perp})e^{i m (\phi-\phi')}\,,
\label{eq:3D_Propagator}
\end{align}
where $J_{m}(x)$ is the $m$th Bessel function, and $k_{nm}\left(2n+1 + |m| +\frac{1}{2}\right) \frac{\pi}{2L}$, $L$ is the length of the system, and ${\bf r}_{\perp} = (r_{\perp}, \phi)$. 

From Eqs.~(\ref{eq:trapped_Propagator}, \ref{eq:3D_Propagator}) one can then show in the limit ${\bf r}, {\bf r}' \to 0$:
\begin{align}
\nabla_{{\bf r}} \cdot &  \nabla_{{\bf r}'} \left[G^0(E,{\bf r},{\bf r}')-G^0_{3D}(E,{\bf r},{\bf r'})\right] = \nonumber \\
&\frac{2\mu}{\pi a_{\perp}^3} \delta_{m,0} \left[\sum_{n=0}^N \sqrt{n- \mathcal{E}_0 - i \delta} -\frac{2}{3}N^{3/2}\right] \nonumber \\
-&\frac{2\mu}{2 \pi a_{\perp}^3} \delta_{|m|,1} \sum_{n=0}^N \left[ \sqrt{n-\mathcal{E}_1- i \delta} - \frac{2}{3}N^{3/2}\right. \nonumber \\
& \left. + \frac{E}{2\omega_{\perp}}\left( \frac{1}{\sqrt{n-\mathcal{E}_1-i \delta}} - 2N^{1/2}\right) \right]\,,
\label{eq:diff_G}
\end{align}
where $N$ is a UV cutoff which should be taken to infinity. The sums in Eq.~(\ref{eq:diff_G}) are regularized and are equivalent to the Hurwitz zeta functions, $\zeta(s,x)$ \cite{zeta_book}. This gives the final result, Eqs.~(\ref{eq:T_0}-\ref{eq:T_1}), for the T-matrix inside the harmonic trapping potential.

The calculation for a q2D system is identical to the q1D case, except that one now uses the q2D basis functions, Eq.~(\ref{eq:q2d_basis}) to construct the propagator in the presence of the trap. The free-space propagator regularizes the leading divergent behavior of the q2D T-matrix, producing Eqs.~(\ref{eq:T0_2D}-\ref{eq:T1_2D}). However, we find that for low-dimensional p-wave interactions, there is still a divergence present that is not regularized by the 3D interaction. This is due to our approximation for the low-dimensional T-matrix following Eq.~\eqref{eq:ansatz}. This additional divergence is identical to the one expected for a true low-dimensional p-wave interaction and can then be regularized by examining the effective low-dimensional scattering at low energies. This additional divergence does not appear if one examines the dimensional crossover using the two-body wavefunction \cite{Zhang17}. However we note that both the T-matrix and wavefunction approaches give consistent results once we fixed this spurious divergence.

\section{Alternative Solution to the T-matrix for q2D}
\label{app:parish}

\subsection{Deriving the Equations for the Binding Energy}

In this section, we consider an identical way of solving the two-body T-matrix following Ref.~\cite{LevinsenParish:2015}. In this method we use a modified pseudo-potential which in 3D has the form:
\begin{equation}
    \langle {\bf k} | v | {\bf k'}\rangle=2\frac{g}{L^2}e^{-\frac{k^2+k'^2}{\Lambda^2}}{\bf k \cdot k'}\,,
\end{equation}
with $\Lambda$ acting as a soft UV cutoff, compared to Eq.~\eqref{eq:pseudopotential} which uses a hard UV cutoff. Solving for the two-body T-matrix in 3D the bare coupling constant $g$ can be related to the 3D scattering parameters:
\begin{equation}
    \frac{1}{g} = \frac{2\mu}{24\pi}\left[ \left(\frac{1}{V_p} - \frac{\Lambda^3}{4\sqrt{2\pi}} \right) + \left(\frac{1}{R_p} - \frac{\Lambda}{\sqrt{2\pi}}\right) 2\mu E\right]\,,
    \label{eq:g_1}
\end{equation}
where $V_p$ and $R_p$ are the 3D $p$-wave scattering volume and effective range, respectively. Once these parameters are set, the physics should be identical to the zero-ranged pseudo-potential used in the main text.

From this modified pseudo-potential, the q2D geometry can be readily evaluated:
\begin{equation}
    \langle {\bf k}, n,m| v | {\bf k'},n',m'\rangle = 2\frac{g}{L^2}f_{n,m}f_{n',m'} ({\bf k \cdot k'})^{1-m} \delta_{m,m'}\,,
\end{equation}
where $m=0,1$. The constants $f_{n,m}$ are related to the transverse wavefunction and its derivative evaluated at the origin $\psi_{n,m}(0)$:
\begin{equation}
    f_{n,m}= \sum_{k_{\perp}}e^{-\frac{k_{\perp}^2}{\Lambda^2}}k_{\perp}^{m}\psi_{n,m}(k_{\perp})\,.
\end{equation}

The benefit of this representation is that $f_{n,m}$ can be evaluated exactly:
\begin{align}
f_{n,m} &= (-1)^n (-i)^m \frac{2^{m/2}}{\pi^{1/4}a_{\perp}^{(2m+1)/2}} \sqrt{\frac{(2n+m)!}{2^{2n}(n!)^2}} \nonumber \\
&\frac{1}{(1+\lambda)^{(2m+1)/2}} \left(\frac{1-\lambda}{1+\lambda}\right)^n\,,
\end{align}
with $\lambda = 2/(\Lambda^2 a^2_{\perp})$. From here the two-body scattering T-matrix following Eq.~\eqref{eq:v_trap} can be written as:
\begin{align}
T_{m}^{-1} = \frac{1}{g} &- \frac{1}{2}\sum_{n=0}^{\infty} \int \frac{d^2p}{(2\pi)^2} \left(\frac{p^{2}}{2}\right)^{1-m} \nonumber \\
&\frac{|f_{n,m}|^2 e^{-2p^2/\Lambda^2}}{E-\frac{p^2}{2\mu} - (2n+|m|+1/2)\omega_{\perp}} \,.\label{app:T_m_parish}
\end{align}
In Eq.~\eqref{app:T_m_parish} $g$ is defined in Eq.~\eqref{eq:g_1}, $\mu$ is the reduced mass and $a_{\perp}^2 = (\mu \omega_{\perp})^{-1}$. When investigating the two-body bound state: \mbox{$E = (|m|+1/2)\omega_{\perp}-\kappa_m^2/(2\mu)$} with binding energy $\kappa_{m}^2/(2\mu)$. In this case, the denominator of Eq.~\eqref{app:T_m_parish} is positive semidefinite, and allows for an analytical summation over the transverse quantum states. Namely, by denoting: $u = p^2/(2\mu)/(\kappa^2_m/(2\mu)+2n\omega_{\perp})$, $\mathcal{E}_m = \kappa_m^2/(4\mu\omega_{\perp})$ and using the relations:
\begin{align}
\sum_{n=0}^{\infty} \frac{(2n)!}{(n!)^2}x^n &= \frac{1}{\sqrt{1-4x}}\,,\nonumber \\ 
\sum_{n=0}^{\infty} \frac{(2n)!}{(n!)^2} n x^n &= \frac{2x}{(1-4x)^{3/2}}\,, \nonumber \\ 
\sum_{n=0}^{\infty} \frac{(2n+1)!}{(n!)^2}x^n &= \frac{1}{(1-4x)^{3/2}}\,, \nonumber \\ 
\end{align}
one can obtain Eqs.~(\ref{eq:2DpwaveDimer}-\ref{eq:2DswaveDimer}).

\subsection{Low-Dimensional Limits of the Binding Energies}

Next we consider the limiting behaviours of Eqs.~(\ref{eq:2DpwaveDimer}-\ref{eq:2DswaveDimer}). In order to discuss the limits, we will focus on the behaviour of the virtual excitation term at small and large binding energies, corresponding to the low- and high-dimensional limits. To that end let us define:
\begin{align}
    J_0 &= \int_0^{\infty} \frac{du}{\lambda} \frac{ue^{-\mathcal{E}_0 u}}{4\lambda + u} \frac{1}{\eta^{3/2}} \left[\mathcal{E}_0 \eta + \frac{1}{2}e^{-u}(1-\lambda)^2\right]\,, \label{eq:J0} \\
    J_1 &= \int_0^{\infty} du \frac{e^{-u}}{4\lambda+u} \frac{1}{\eta^{3/2}}\,, \label{eq:J1}
\end{align}
where $\eta =(1+\lambda)^2-e^{-u} (1-\lambda)^2$.

In the low-dimensional limit, $\mathcal{E}_{0,1} \ll 1$, we expect that the regularization terms are not important as they regulate the high energy, or 3D, physics. In this case we can simplify the integral by noting that the virtual excitation term for small binding energies is weighted at large $u$, and as such: $\eta \sim 1$. Restricting the region of integration to $u>4\lambda$ allows one to evaluate the integral analytically. Then taking the limit $\lambda \to 0$ we find:
\begin{align}
    J_0 &\approx  \frac{1}{\lambda} -4\mathcal{E}_0\ln\left(\frac{e^{-\gamma_E}}{4\lambda \mathcal{E}_0}\right)\,, \\
    J_1 &\approx \ln\left(\frac{e^{-\gamma_E}}{4\lambda \mathcal{E}_1}\right)\,.
\end{align}
We note that the q2D p-wave term $J_0$ has an additional divergence of $\lambda^{-1}$. As discussed previously, this is exactly the divergence one would expect for a true 2D p-wave system, and is an artifact of our solution to the T-matrix. As such, this term is required to regularize the q2D T-matrix, and can be safely removed. The resulting terms are the correct energy dependencies for true 2D $p$-wave and $s$-wave systems respectively. Substituting these expressions into Eqs.~(\ref{eq:2DpwaveDimer}-\ref{eq:2DswaveDimer}) yield the correct low-dimensional equations for the binding energies. 

In the opposite limit $\mathcal{E}_{0,1} \gg 1$ we expect to recover the 3D equation for the binding energy. In this case, we expect the virtual excitation term to produces divergences that will cancel the regularization terms in Eqs.~(\ref{eq:2DpwaveDimer}-\ref{eq:2DswaveDimer}), i.e.\ terms that scale as $\lambda^{-3/2}$ and $\lambda^{-1/2}$, while the leading regular term scales as $\mathcal{E}_{0,1}^{3/2}.$  In this limit the dominant contribution in Eqs.~(\ref{eq:J0}-\ref{eq:J1}) is at small $u$. For the $s$-wave case, one can approximate $\eta \approx u$ to find 
\begin{equation}
    J_1 \approx \frac{1}{12} \frac{1}{\lambda^{3/2}} - \frac{\mathcal{E}_1}{\lambda^{1/2}} + \frac{4\sqrt{\pi}}{3} \mathcal{E}_1^{3/2}
\end{equation}
which gives the correct 3D result when substituted into Eq.~\eqref{eq:2DswaveDimer}. An analogous analytic expression for the limiting case of $J_0$ was not found, however the full numerical solutions of Eqs.~(\ref{eq:2DpwaveDimer}-\ref{eq:2DswaveDimer}) shown in Fig.~\ref{fig:q2D_BE} confirm that the bound states do indeed approach the 3D solution of Eq.~\eqref{eq:3D_BS}. 

\section{Two-Body RF and Operator Product Expansion}
\label{app:ope_appendix}
Consider an initial state $|Q, N, M, k,n,m\rangle$ with $Q,N,M$ being the center of mass momentum, principal transverse quantum number, and the transverse angular momentum (or parity), respectively. Similarly $k,n,m$ correspond to the quantum numbers for the relative momentum, principal transverse quantum number, and the transverse angular momentum (or parity). The main contribution to the RF spectroscopy at the two-body level comes from Fig.~\ref{fig:OPE}, where the black circles are the two-body T-matrices in Eq.~(\ref{eq:T_0}-\ref{eq:T_1}), and the open circles are a) the operator insertions from Eq.~(\ref{eq:RF}) or b) the contact operators. This is equally valid for both q1D and q2D. For simplicity of the presentation, we will ignore the center of mass coordinate and provide explicit calculations for the case of q1D. The extension to q2D is straightforward with minimal change. 

The first step in our OPE calculation is to evaluate Eq.~(\ref{eq:RF}) using our two-body state. For our chosen initial state and q1D scattering, the diagram in Fig.~\ref{fig:OPE}(a) has the form:
\begin{align}
I_m(\omega) &\approx -\sum_{n'} \int \frac{d\ell}{2\pi} \left[f_{n}^{(m)}(k) T^{(m)}(E)\right]^2 \left(f_{n'}^{(m)}(\ell)\right)^2 \nonumber \\
&\times \frac{1}{E+\omega - E_{n',m}(k) + i \delta}\frac{1}{\left(E- E_{n',m}(k) + i \delta\right)^2}\,,
\label{eq:2body}
\end{align}
where we have performed the internal frequency integral, and we have also defined the form factors: $f_{n}^{(0)}(k) = k \sqrt{1 /(\pi a_{\perp}^2)}$ and $f_n^{(\pm 1)} = \sqrt{2 /(\pi a_{\perp}^4)}$. Here we also  use $T^{(m)}$ are the solutions to the Lipmann-Schwinger equation with definite angular momentum (parity) $m$.

The high-frequency tails of the RF transfer rate arise from the  non-analytic terms in Eq.~(\ref{eq:2body}). From the non-analytic terms in Eq.~(\ref{eq:2body}), one finds the two-body solution for the RF transfer rate:
\begin{align}
\Gamma^{(0)}&\approx \left[f_{n}^{(0)}(k) T^{(0)}(E)\right]^2  \nonumber \\
&\times\sum_n \frac{1}{8\pi} \frac{(2\omega_{\perp})^{3/2}}{\omega^2} \sqrt{\mathcal{E}_0-n} \ \theta(\mathcal{E}_0 -n)\,,
\label{eq:2body_odd}
\end{align}
for $p$-wave interactions $(m=0)$, and for $s$-wave interactions $(m=\pm 1)$:
\begin{align}
\Gamma^{(\pm1)}&\approx \left[f_{n}^{(\pm1)}(k) T^{(\pm1)}(E)\right]^2  \nonumber \\
&\times\sum_n \frac{n+1}{16\pi} \frac{(2\omega_{\perp})^{3/2}}{\omega^2} \frac{1}{\sqrt{\mathcal{E}_{1}-n}} \theta(\mathcal{E}_1 -n)\,,
\label{eq:2body_even}
\end{align}
where again  $\mathcal{E}_m = (E - (|m|+1) \omega_{\perp})/(2\omega_{\perp})$.

Eqs.~(\ref{eq:2body_odd}-\ref{eq:2body_even}) are the same as Eqs.~(\ref{eq:RF_2B_odd_1D}-\ref{eq:RF_2B_even_1D}), where we have replaced the factor that depends on the initial state, $\left[f_{n}^{(m)}(k) T^{(m)}(E)\right]^2$, with a constant that depends on the many-body physics, $A_m(E_F,T)$.

The second step of the OPE is to connect this two-body calculation to the expectation value of local operators. The operators of interest to us are the 3D contact operators discussed in Appendix \ref{appendix:review}. To facilitate our calculation, we can express the contact operators in the basis of two-body states in the presence of transverse confinement. It is straightforward to show the contact operators have the form:
\begin{align}
C_{V_p} =\sum_m  C_{V_{p}}^{(m)} &=\sum_m \sum_{Q,N,M} \sum_{k,n} \sum_{k',n'} \nonumber \\
&\frac{g^2}{12\pi}f_n^{(m)}(k) f_{n'}^{(m)}(k')\left(\Phi_{Q,N,M}^{k,n,m}\right)^{\dagger}\Phi_{Q,N,M}^{k',n',m}\,, \nonumber \\
C_{R_{p}} =\sum_m  C_{R_{p}}^{(m)} &=\sum_m \sum_{Q,N,M} \sum_{k,n} \sum_{k',n'} \nonumber \\
&\left(E-\frac{Q^2}{4}-(2N+|M|+1)\omega_{\perp}\right)\nonumber \\
&\frac{g^2}{12\pi}f_n^{(m)}(k) f_{n'}^{(m)}(k')\left(\Phi_{Q,N,M}^{k,n,m}\right)^{\dagger}\Phi_{Q,N,M}^{k',n',m}\,,
\label{eq:contact_operators}
\end{align}
where $g$ is the $p$-wave coupling strength, $f_n^{(0)}(k) = \sqrt{1/\pi a_{\perp}^2}k$ and $f_n^{(\pm 1)} = \sqrt{2(n+1)/\pi a_{\perp}^4}$ are the q1D form factors. The operator  $\Phi_{Q,N,M}^{k,n,m}$ is the annihilation operator for a two-body state with energy $E$, center of mass quantum numbers $Q,N,M$, and relative quantum numbers $k,n,m$.

At the two-body level, the calculation of the expectation value of the contact operators are related to the diagrams in Fig.~\ref{fig:OPE})(b). The result is:
\begin{align}
\langle C_{V_p} \rangle  &= \sum_m \frac{2}{12\pi} \left[f_{n}^{(m)}(k) T^{(m)}(E)\right]^2\,, \nonumber \\
\langle C_{R_p} \rangle  &= \sum_m \frac{2}{12\pi} \left[f_{n}^{(m)}(k) T^{(m)}(E)\right]^2 E\,. \nonumber \\
\label{eq:RHS}
\end{align}

Eq.~(\ref{eq:RHS}) matches the leading and subleading energy dependence of Eqs.~(\ref{eq:2body_odd}-\ref{eq:2body_even}). From there it is possible to extract the Wilson coefficients, $W(\omega)$, for the scattering volume and effective range contacts. This is most readily done in the 1D limit, $\omega \ll 2\omega_{\perp}$, and in the 3D limit, $\omega_{\perp} \ll \omega$. In the 1D limit the Wilson coefficients depend on the angular momentum, $m$:
\begin{align}
&C_{V_p}^{(0)}: & W(\omega) &=  \frac{\Omega^2}{2\omega^{3/2}} \frac{6}{a_{\perp}^2} \left(1- \frac{1}{a_{\perp}^2\omega}\right)\,,  \nonumber \\
&C_{V_p}^{(\pm 1)}: &  W(\omega) &= \frac{\Omega^2}{4\omega^{5/2}} \frac{24}{a_{\perp}^4}\,, \nonumber \\
&C_{R_p}^{(0)}: & W(\omega) &= \frac{\Omega^2}{2\omega^{5/2}} \frac{3}{a_{\perp}^2}\,,
\label{eq:1d_wilson}
\end{align}
Conversely, in the 3D limit, the Wilson coefficients are isotropic:
\begin{align}
&C_{V_{p}}^{(m)}: & W(\omega)&= \frac{\Omega^2}{2\omega^{1/2}} \,,\nonumber \\
&C_{R_{p}}^{(m)}: & W(\omega)&= \frac{3\Omega^2}{4\omega^{3/2}}\,,
\label{eq:3d:wilson}
\end{align}
as expected in the true 3D limit. The Wilson coefficients in Eqs.~(\ref{eq:1d_wilson}-\ref{eq:3d:wilson}) produce the 1D and 3D results.

The calculation described above can be readily extended to the case of q2D scattering. There one finds the following Wilson coefficients in the 2D limit ($\omega \ll 2\omega_{\perp}$):
\begin{align}
&C_{V_p}^{(0)}: & W(\omega) &=  \frac{\pi\Omega^2}{2\omega} \frac{3}{2\sqrt{\pi}a_{\perp}} \left(1-\frac{1}{a_{\perp}^2\omega}\right) \,, \nonumber \\
&C_{V_p}^{(1)}: &  W(\omega) &=  \frac{\pi \Omega^2}{4\omega^{2}} \frac{12 }{\sqrt{\pi} a_{\perp}^3}\,, \nonumber \\
&C_{R_p}^{(0)}: & W(\omega) &= \frac{\pi \Omega^2}{2\omega^2} \frac{3}{2 \sqrt{\pi} a_{\perp}} \,.
\label{eq:2d_wilson}
\end{align}
In the 3D limit ($\omega \gg \omega_{\perp}$), the Wilson coefficients are the same as those in Eq.~(\ref{eq:3d:wilson}), recovering the true 3D result.

\section{Review of d-Dimensional Contacts and RF Spectroscopy}
\label{appendix:review}
In this section we review the two-body scattering for $d$-dimensional gases. We also quote the relevant definitions for the contact operators, and the RF transfer function. For a $d$-dimensional gas with $s$- and $p$-wave interactions we consider the following Hamiltonians, $H_s$ and $H_p$, respectively:
\begin{align}
H_s &= \int d^d{\bf r} \psi_{\sigma}^{\dagger}({\bf r}) \left(-\frac{1}{2}\nabla_{\bf r}^2 \right) \psi_{\sigma}({\bf r}) \nonumber \\
&+ g \int d^d{\bf r} \psi_{\uparrow}^{\dagger}({\bf r})\psi_{\downarrow}^{\dagger}({\bf r}) \psi_{\downarrow}({\bf r})\psi_{\uparrow}({\bf r})\,, \label{eq:Hs} \\
H_p &= \int d^d{\bf r} \psi^{\dagger}({\bf r}) \left(-\frac{1}{2}\nabla_{\bf r}^2 \right) \psi({\bf r}) \nonumber \\
&- \frac{g}{2} \int d^d{\bf r} \psi^{\dagger}({\bf r})\overleftrightarrow{\boldsymbol{\nabla}}_{\bf r} \psi^{\dagger}({\bf r}) \psi({\bf r})\overleftrightarrow{\boldsymbol{\nabla}}_{\bf r} \psi({\bf r})\,.
\label{eq:Hp}
\end{align}
In defining $H_s$ and $H_p$ we let $\psi_{(\sigma)}$ is the annihilation operator for the (spin-$1/2$) spin-polarized fermions, $g$ is the ($s$) $p$-wave interaction strength, and we have defined: $\overleftrightarrow{\boldsymbol{\nabla}}_{\bf r} = (\overrightarrow{\boldsymbol{\nabla}}_{\bf r}-\overleftarrow{\boldsymbol{\nabla}}_{\bf r})/2$. We have also set $\hbar = m = 1$. These Hamiltonians are consistent with the pseudo-potential shown in Eq.~(\ref{eq:pseudopotential}).

The low-dimensional T-matrices are defined in Eqs.~(\ref{eq:T_1D_p}, \ref{eq:T_1D_s}, \ref{eq:T_2D_p}, \ref{eq:T_2D_s}), which depends on the low-dimensional scattering parameters: $a_{s,d\mathrm{D}}$ the $s$-wave scattering length and $r_{s,d\mathrm{D}}$ the $s$-wave effective range, $a_{d\mathrm{D},p}$, the $p$-wave scattering volume, and $r_{d\mathrm{D},p}$ the $p$-wave effective range.

\subsection{Thermodynamic Contacts}
The low-dimensional contacts are defined in terms of the change of energy with respect to the low-dimensional scattering parameters. For $s$-wave interactions the contacts are:
\begin{align}
    d&= 1 & C_{a_s} &= \frac{\partial H_{s}}{\partial a_{s}} & C_{r_s} &= -\frac{\partial H_s}{\partial r_s} \label{eq:def_C_1Ds}\\
    d&= 2 & C_{a_s} &= -\frac{\partial H_{s}}{\partial \ln\left(a_s^{-2}\right)} & C_{r_s} &= -\frac{\partial H_s}{\partial r_s}
    \label{eq:def_C_2Ds}
\end{align}
while for $p$-wave interactions \cite{Cui16a, Cui16b}:
\begin{align}
    d&= 1 & C_{a_p} &= -\frac{\partial H_{p}}{\partial a_{p}^{-1}} & C_{r_p} &= -\frac{\partial H_p}{\partial r_p} \label{eq:def_C_1Dp}\\
    d&= 2 & C_{a_p} &= -\frac{\partial H_{p}}{a_p^{-1}} & C_{r_p} &= -\frac{\partial H_p}{\partial \ln\left(r_p^{-2}\right)} \label{eq:def_C_2Dp}\\
    d&= 3 & C_{a_p} &= -\frac{\partial H_{p}}{\partial V_p^{-1}} & C_{r_p} &= -\frac{\partial H_s}{\partial R_p^{-1}} \label{eq:def_C_3Dp}
\end{align}
In Eqs.~(\ref{eq:def_C_1Ds}-\ref{eq:def_C_3Dp}) we have muted the dimensional index on the scattering parameters for transparency. Microscopically, the contacts in $d$ dimensions can be succinctly written as:
\begin{align}
C_{a_s} &= \frac{g^2}{\mathcal{A}_s} \int d^d{\bf r} \psi_{\uparrow}^{\dagger}({\bf r})\psi_{\downarrow}^{\dagger}({\bf r}) \psi_{\downarrow}({\bf r})\psi_{\uparrow}({\bf r})\,, \label{eq:Ca_s} \\
C_{r_s} &= \frac{g^2}{\mathcal{A}_s} \int d^d{\bf r} \psi_{\uparrow}^{\dagger}({\bf r})\psi_{\downarrow}^{\dagger}({\bf r})\left[i \partial_t - H_\mathrm{cm}\right]\psi_{\downarrow}({\bf r})\psi_{\uparrow}({\bf r})\,, \label{eq:Cr_s} \\
C_{a_p} &= -\frac{g^2}{\mathcal{A}_p} \int d^d{\bf r} \frac{1}{2}\psi^{\dagger}({\bf r})\overleftrightarrow{\boldsymbol{\nabla}}_{\bf r} \psi^{\dagger}({\bf r}) \psi({\bf r})\overleftrightarrow{\boldsymbol{\nabla}}_{\bf r} \psi({\bf r})\,, \label{eq:Ca_p}\\
C_{r_p} &=  -\frac{g^2}{\mathcal{A}_p} \int d^d{\bf r} \frac{1}{2}\psi^{\dagger}({\bf r})\overleftrightarrow{\boldsymbol{\nabla}}_{\bf r} \psi^{\dagger}({\bf r})\left[i \partial_t - H_\mathrm{cm} \right]\psi({\bf r})\overleftrightarrow{\boldsymbol{\nabla}}_{\bf r} \psi({\bf r})\,,
\label{eq:Cr_p}
\end{align}
where $H_\mathrm{cm}$ is the center of mass momentum. The explicit definitions of the contacts and the values of $\mathcal{A}_s$ and $\mathcal{A}_p$ are tabulated in Tab.~\ref{tab:coefficients}. 

\begin{table}[]
\begin{tabular}{c@{\hskip 0.3in}c@{\hskip 0.3in}c}
\hline
$d$ & $\mathcal{A}_{s}$ & $\mathcal{A}_{p}$ \\ \hline
$1$ & $2$                 & $2$                 \\ \hline
$2$ & 4$\pi$              & $8\pi$             \\ \hline
$3$ & $4\pi$              & $12\pi$             \\ \hline
\end{tabular}
\caption{Parameters $\mathcal{A}_{s}$ and $\mathcal{A}_{p}$ used in defining the microscopic formulae for the contacts, Eqs.~(\ref{eq:Ca_s}-\ref{eq:Cr_p}).}
\label{tab:coefficients}
\end{table}

\subsection{RF Spectroscopy}
The RF transfer rate is microscopically defined in Eq.~(\ref{eq:RF}). The large detuning tails of the RF transfer rate can be obtained using the operator product expansion (OPE) technique \cite{Braaten10}. Here we simply quote the relevant results for $s$-wave interactions \cite{Braaten10,Langmack12}:
\begin{align}
    d&= 1 & \Gamma_s(\omega) &= \frac{\Omega^2}{4\omega^{5/2}} \langle C_{a_s} \rangle \label{eq:Gamma_1D_s}\\
    d&= 2 & \Gamma_s(\omega) &= \frac{\pi\Omega^2}{4\omega^{2}} \langle C_{a_s} \rangle \label{eq:Gamma_2D_s}
\end{align}
and for $p$-wave interactions \cite{Cui16a, Zhang17, Luciuk16}:
\begin{align}
    d&= 1 & \Gamma_{p}(\omega) &= \frac{\Omega^2}{2\omega^{3/2}} \left[\langle C_{a_{p}} \rangle + \frac{1}{2\omega} \langle C_{r_{p}} \rangle\right] \label{eq:Gamma_1D_p}\\
    d&= 2 & \Gamma_{p}(\omega) &= \frac{\pi \Omega^2}{2\omega} \left[\langle C_{a_{p}} \rangle + \frac{1}{\omega} \langle C_{r_{p}} \rangle\right] \label{eq:Gamma_2D_p}\\
    d&= 3 & \Gamma_p(\omega) &= \frac{\Omega^2}{2\omega^{1/2}} \left[\langle C_{a_p} \rangle + \frac{3}{2\omega} \langle C_{r_p} \rangle\right]
\end{align}

\bibliography{theory} 

\end{document}